\pdfoutput=1
\documentclass[aps,prd,amsmath,floats,floatfix, twocolumn,
superscriptaddress,nofootinbib,showpacs,longbibliography]{revtex4-1}

\usepackage[T1]{fontenc}
\usepackage[utf8]{inputenc}
\usepackage{lmodern}
\usepackage{times}
\usepackage[varg]{txfonts}
\usepackage{verbatim}

\usepackage[dvipsnames, usenames]{xcolor}
\definecolor{linkcolor}{rgb}{0.0,0.3,0.5}
\usepackage[hypertexnames=false, unicode, colorlinks=true, linkcolor=linkcolor,
citecolor=linkcolor, filecolor=linkcolor,urlcolor=linkcolor,
pdfusetitle]{hyperref}

\usepackage[all]{hypcap}
\usepackage{graphicx}
\usepackage{xspace}
\usepackage{amssymb}
\usepackage[normalem]{ulem} %
\usepackage{bm} %

\usepackage{microtype}

\usepackage[english]{babel}
\usepackage{blindtext}
\usepackage{orcidlink}
\graphicspath{%
  {figs/}%
}

\DeclareMathAlphabet{\mathpzc}{OT1}{pzc}{m}{it}

\newcommand{\into}{\!\times\!\relax} %

\newcommand{\h}{\mathpzc{h}}
\newcommand{\hplus}{h_{+}}
\newcommand{\hcross}{h_{\into}}
\newcommand{\hlm}{\mathpzc{h}_{\ell m}}
\newcommand{\htwotwo}{\mathpzc{h}_{22}}

\newcommand{\dd}{\mathrm{d}}

\newcommand{\package}{\texttt{gw\_eccentricity}\xspace}
\newcommand{\egw}{e_{\text{gw}}}
\newcommand{\etwotwo}{e_{\omega_{22}}}

\newcommand{\omegaA}{\omega^{\text{a}}_{22}}
\newcommand{\omegaP}{\omega^{\text{p}}_{22}}

\newcommand{\omegatwotwo}{\omega_{22}}

\newcommand{\avgOmega}{\langle\omega_{22}\rangle}

\newcommand{\avgT}{\langle t\rangle}
\newcommand{\avgTP}{\langle t\rangle^{\text{p}}}
\newcommand{\avgTA}{\langle t\rangle^{\text{a}}}
\newcommand{\phitwotwo}{\phi_{22}}
\newcommand{\eEOB}{e_{\text{eob}}}
\newcommand{\eGeneric}{e\xspace}
\newcommand{\lGeneric}{l\xspace}

\newcommand{\tref}{t_{\text{ref}}}
\newcommand{\fref}{f_{\text{ref}}}
\newcommand{\avgfref}{\langle f_{\text{ref}}\rangle}

\newcommand{\mResAmp}{\texttt{ResidualAmplitude}\xspace}

\newcommand{\Atwotwo}{A_{22}}

\newcommand{\SEOB}{\texttt{SEOBNRv4EHM}}

\newcommand{\SEOBNRE}{\texttt{SEOBNRE}\xspace}
\newcommand{\TEOB}{\texttt{TEOBResumS-Dal\'i}\xspace}
\newcommand{\TEOBGeneric}{\texttt{TEOBResumS}\xspace}
\newcommand{\EccentricTD}{\texttt{EccentricTD}\xspace}

\newcommand{\tX}{t^{\text{X}}}
\newcommand{\tP}{t^{\text{p}}}
\newcommand{\tA}{t^{\text{a}}}
\newcommand{\dm}{(2, 2)\xspace}
\newcommand{\final}{f\xspace}
\newcommand{\mdev}{\frac{\Delta M_{\final}}{\bar{M}_{\final}}\xspace}
\newcommand{\mdevinline}{\Delta M_{\final}/\bar{M}_{\final}\xspace}
\newcommand{\sdev}{\frac{\Delta \chi_{\final}}{\bar{\chi}_{\final}}\xspace}
\newcommand{\sdevinline}{\Delta \chi_{\final}/\bar{\chi}_{\final}\xspace}
\newcommand{\spinvec}{\mathbf{S}\xspace}
\newcommand{\spin}{\chi\xspace}
\newcommand{\spinpair}{(\spin_1, \spin_2)\xspace}

\newcommand{\insp}{I\xspace}
\newcommand{\postinsp}{MR\xspace}
\newcommand{\spinfInsp}{\chi_{\final}^{\text{\insp}}}
\newcommand{\spinfPostInsp}{\chi_{\final}^{\text{\postinsp}}}
\newcommand{\massfInsp}{M_{\final}^{\text{\insp}}}
\newcommand{\massfPostInsp}{M_{\final}^{\text{\postinsp}}}
\newcommand{\mchirp}{m_1^{3/5}m_2^{3/5}/(m_1 + m_2)^{1/5}\xspace}
\newcommand{\Mc}{\mathcal{M}_c\xspace}
\newcommand{\QGR}{Q_{\texttt{GR}}\xspace}
\newcommand{\bilby}{\texttt{Bilby}~\cite{Ashton:2018jfp}\xspace}
\newcommand{\qc}{quasicircular\xspace}
\newcommand{\Lvec}{\mathbf{L}\xspace}
\newcommand{\tilt}{\theta^{\spinvec\Lvec}\xspace}

\begin{document}

\title{A study of the Inspiral-Merger-Ringdown Consistency Test with gravitational-wave \\ signals from compact binaries in eccentric orbits}

\newcommand{\ICTS}{\affiliation{International Centre for Theoretical Sciences,
    Tata Institute of Fundamental Research, Bangalore 560089, India}}
\newcommand{\IUCAA}{\affiliation{The Inter-University Centre for Astronomy and
    Astrophysics, Post Bag 4, Ganeshkhind, Pune 411007, India}}
\newcommand{\SNU}{\affiliation{Department of Physics and Astronomy,
    Seoul National University, Seoul 08826, Korea}}
\newcommand{\VSM}{\affiliation{Department of Physics, Vivekananda Satavarshiki Mahavidyalaya (affiliated to Vidyasagar University), Manikpara 721513, West Bengal, India}}
\newcommand{\CMI}{\affiliation{Chennai Mathematical Institute, Plot H1 SIPCOT IT Park, Siruseri 603103, India.}}

\author{Md Arif Shaikh\,\orcidlink{0000-0003-0826-6164}}
\email{arifshaikh.astro@gmail.com}
\VSM
\SNU
\ICTS
\author{Sajad A. Bhat}
\email{sajad.bhat@iucaa.in}
\thanks{The first two authors contributed equally}
\IUCAA
\CMI
\author{Shasvath J. Kapadia\,\orcidlink{0000-0001-5318-1253}}
\email{shasvath.kapadia@iucaa.in}
\IUCAA
\hypersetup{pdfauthor={Shaikh et al.}}

\date{\today}

\begin{abstract}
The Inspiral Merger Ringdown Consistency Test (IMRCT) is one among a battery of
tests of general relativity (GR) employed by the LIGO-Virgo-KAGRA (LVK)
collaboration. It is used to search for deviations from GR in detected
gravitational waves (GWs) from compact binary coalescences (CBCs) in a
model-agnostic way. The test compares source parameter estimates extracted
independently from the inspiral and post-inspiral portions of the CBC signals
and, therefore, crucially relies on the accurate modeling of the
waveform. Current implementations of the IMRCT routinely use \qc waveforms,
under the assumption that the residual eccentricity of the binary when the
emitted GWs enter the frequency band of the LVK detector network will be
negligible. In this work, we perform a detailed study to investigate the
typical magnitudes of this residual eccentricity that could potentially lead to
spurious violations of the IMRCT. To that end, we conduct injection campaigns
for a range of eccentricities and recover with both \qc and eccentric
waveforms. We find that an eccentric GW signal from a GW150914-like system with
eccentricity $\egw \gtrsim$ 0.04 at an orbit averaged frequency $\langle
f_{\text{ref}} \rangle=$ 25 Hz breaks the IMRCT if recovered with \qc waveforms
at $\gtrsim 68\%$ confidence. The violation becomes more severe ($\gtrsim 90\%$
confidence) for $\egw =$ 0.055 at $\langle f_{\text{ref}} \rangle=$ 25 Hz. On
the other hand, when eccentric waveforms are used, the IMRCT remains intact for
all eccentricities considered. We also briefly investigate the effect of the
magnitude and orientation (aligned/antialigned) of the component spins of the
binary on the extent of the spurious violations of the IMRCT. Our work,
therefore, demonstrates the need for accurate eccentric waveform models in the
context of tests of GR.
\end{abstract}

\maketitle

\section{Introduction}
\label{sec:introduction}
The LIGO-Virgo~\cite{TheLIGOScientific:2014jea, TheVirgo:2014hva} network of ground-based interferometric gravitational-wave (GW)
detectors has detected $\sim 90$ compact binary coalescence (CBC) events in its
first three observing runs (O1, O2, O3)~\cite{KAGRA:2021vkt}. The majority of
these are binary black hole (BBH) mergers, although binary neutron star
(BNS)~\cite{TheLIGOScientific:2017qsa, Abbott:2020uma} and neutron star-black
hole (NSBH) coalescences~\cite{LIGOScientific:2021qlt} have also been observed.

The detected GW signals, especially those pertaining to BBH mergers, come from
the last few orbits of the inspiral, as well as the merger and ringdown phases
(see, e.g.,~\cite{Sathyaprakash:2009xs}). These GWs are well-suited to probe
general relativity (GR) in the strong-field regime, unlike most tests of GR
that use electromagnetic waves, which typically probe the weak field
regime\footnote{A notable exception to this are the results of the Event Horizon
Telescope (EHT)~\cite{EventHorizonTelescope:2019dse}, which imaged the shadow
of the supermassive black hole at the center of the M31 galaxy and enabled electromagnetic-based strong-field tests of GR~\cite{Psaltis:2018xkc}.}.

The LIGO-Virgo-KAGRA (LVK) collaboration ~\cite{TheLIGOScientific:2014jea,
TheVirgo:2014hva, KAGRA:2020tym} has conducted a suite of tests of GR across
O1, O2, and O3 \cite{LIGOScientific:2021sio}. This includes\footnote{but is not
  limited to}

\begin{itemize}
\item A model-agnostic residuals test, which subtracts out the
  best-matched GR-modeled CBC waveform from the data containing the GW signal,
  and checks if the statistical properties of the residual are consistent with
  noise \cite{LIGOScientific:2016lio, LIGOScientific:2019fpa,
    LIGOScientific:2020ufj}
\item A test that probes the inspiral evolution of the CBC
  by searching for deviations in Post-Newtonian (PN) parameters of the GW signal
  \cite{Blanchet:1994ez, blanchet1994signal, Arun:2006hn, Arun:2006yw,
    Yunes:2009ke, Mishra:2010tp, Li:2011cg, Li:2011vx}
\item An inspiral-merger-ringdown
  consistency test (IMRCT) that compares and assesses the consistency between the
  low and high-frequency portions of the CBC signals \cite{Ghosh:2016qgn,
    Ghosh:2017gfp}
\item Propagation tests of GWs that compare their speed with
  respect to the speed of light \cite{LIGOScientific:2018dkp} as well as any
  modulations in the waveform due to finite-graviton-mass-driven velocity
  dispersion \cite{Will:1997bb}.
\end{itemize}

In this work, we restrict our attention to the IMRCT. The IMRCT compares the
two-dimensional joint posterior on the final mass and spin of the merged binary
estimated independently from the inspiral (low-frequency) and post-inspiral
(high-frequency) portions of the waveform \cite{Ghosh:2016qgn,
Ghosh:2017gfp}. A sufficiently large deviation from GR should result in the
difference distribution between final mass and final spin
evaluated from these 2D posteriors to deviate from the origin (zero). However,
it is conceivable that waveforms that do not accurately capture the physics of
the CBC could lead to biased posteriors. We propose to study the resulting
systematics in the IMRCT. In particular, we investigate how the neglect of
eccentricity in the waveform models could cause spurious violations of this
test.

It is well-known that, as GWs carry away energy and angular momentum of a CBC,
it reduces the binary orbit's eccentricity \cite{Peters:1963ux,
Peters:1964zz}. The prevalent expectation is that by the time GWs from the CBC
enter the frequency band of the LVK network, any initial eccentricity that the
binary may have had at the time of formation would have been reduced to a
negligibly small value. Indeed, current constraints on the rate of eccentric
mergers, from searches for such signals, suggest that observations of these in
O4 are likely to be relatively rare \cite{LIGOScientific:2023lpe,
LIGOScientific:2019dag, Nitz:2019spj}. Thus, tests of GR conducted so far by
the LVK routinely use \qc waveforms, including the IMRCT.

On the other hand, certain formation channels, such as those pertaining to
dense stellar environments, could produce binaries whose eccentricities are
non-negligible when they enter the LVK frequency
band~\cite{Mapelli:2021for}. Dynamical encounters in such environments could
harden the binary so that GW emission cannot completely carry away the
eccentricity by the time the GWs enter the LVK's frequency
band~\cite{Samsing:2017xmd, Zevin:2018kzq, Romero-Shaw:2022xko}. Moreover, in
active-galactic-nuclei disks, eccentricity can be amplified by binary-single encounters
\cite{Samsing:2020tda, Tagawa:2020jnc}, while the presence of a third object in
the vicinity of the binary could similarly boost the eccentricity via the
Kozai-Lidov mechanism \cite{Kozai:1962zz, Lidov:1962wjn, Naoz:2016tri,
Antonini:2017ash, Randall:2017jop, Yu:2020iqj, Bartos:2023lfu}. Such
eccentricity-enhancing processes could also leave a residual in-band
eccentricity detectable by the LVK network. Neglecting eccentricity of such
signals in (PN) parametrized tests has already been shown to lead to spurious
violations of GR both at the individual event level~\cite{Saini:2022igm} and
population level inference~\cite{Saini:2023rto}.  We here focus only on the
effect of neglect of eccentricity on the IMRCT.

Performing the IMRCT using \qc waveform models, could potentially cause
systematics in the IMRCT. Intuitively, the eccentricity at the time of
band-entry of the GW signal will be larger than the eccentricity in the
post-inspiral phase, where it is usually vanishingly small. Thus, one would
expect source parameters recovered with \qc waveform models to be biased when
using the inspiral portion of the
waveform~\cite{Favata:2013rwa,Favata:2021vhw,OShea:2021faf}, while those
recovered using the post-inspiral phase to be relatively less biased, if at
all. The resulting 2D difference-distribution constructed from the
(inspiral/post-inspiral) final mass and spin posteriors should be deviated from
zero.

To assess the impact of neglecting eccentricity in waveform models on the
IMRCT, we consider a set of synthetic eccentric GW signals. We inject these in
the most probable realization of (zero-mean) Gaussian noise, the so-called
``zero-noise'' realization\footnote{Here zero-noise realization
refers to the most-probable realization of Gaussian noise with zero mean and
the parameter uncertainties obtained using this noise-realization provide lower
bound to the estimates obtained using the non-zero real noise-realization.},
assuming an O4-like noise power-spectral density
(PSD)~\cite{KAGRA:2013rdx}. We then perform the IMRCT on each of these
signals, using \qc templates. We find that the IMRCT is indeed violated with
\qc templates, {as we increase the injected eccentricity, for a set of masses
and spins where the IMRCT is known not to suffer from other systematics. This
violation progressively increases, from moderate to severe, with increasing
eccentricity. To ascertain that no intrinsic waveform systematics are
contributing to this bias, we repeat the IMRCT using eccentric waveform model
for recovery. We indeed find that the IMRCT is no longer violated.

Past work, using an approximate Fisher Matrix Analysis (FMA)
\cite{Cutler:1994ys} and the Cutler-Vallisneri framework \cite{Cutler:2007mi},
pointed out that the IMRCT could be violated due to neglect of eccentricity
\cite{Bhat:2022amc}. Another work also found spurious violations of the IMRCT
using available numerical-relativity waveforms as injections, while performing
the IMRCT with \qc phenomenological waveform models \cite{Narayan:2023vhm}. Our
work complements and expands on both of these studies in the following way. We
perform Bayesian parameter inference runs using large-scale nested sampler and,
therefore, have more realistic estimate of the bias compared to those in
provided in \cite{Bhat:2022amc}. Moreover, unlike in~\cite{Bhat:2022amc}, we do
not neglect the eccentricity in the post-inspiral part of the signal which may
lead to over-estimation or under-estimation of the bias. Also, because we
employ an eccentric waveform model \TEOB~\cite{Nagar:2018zoe,Nagar:2021gss}, we
have two advantages over the results in ~\cite{Narayan:2023vhm}. We use the
same waveform model to create the injection and then to recover it. Therefore,
we rule out any bias originating due to the waveform systematics which may be
present if the recovery template is not as accurate as the injected NR
waveforms. Also, we have the freedom to create injections at arbitrary values
of eccentricities allowing us to do an injection study on a finer grid of
eccentricity values. This, in turn, allows us to identify more accurately at
what eccentricity the IMRCT bias becomes severe. Similarly, we examine the
sensitivity of the bias for a given eccentricity on other parameters like the
magnitude of the component spins and their orientation (aligned/antialigned)
with respect to the orbital angular momentum of the binary.  Moreover, we
ascertain that the IMRCT is indeed recovered as expected, when eccentricity is
included, for all eccentricities considered. We make contact with both these
works to compare the eccentricity at which the IMRCT is violated, using a
generalised definition of eccentricity that is independent of waveform models
\cite{Shaikh:2023ypz}.

The rest of this paper is organized as follows. Section \ref{sec:methods}
briefly describes the IMRCT, and argues why one should expect spurious
violations due to neglect of eccentricity. Section \ref{sec:results} shares the
results of the injection campaign described above. Section \ref{sec:conclusion}
summarizes the work and discusses other potential sources of spurious IMRCT
violations.

\section{Methods}
\label{sec:methods}
\subsection{Gravitational-wave parameter estimation}
In GR, GWs have two polarizations, $h_{+}$ and $h_{\times}$. The phase
evolution of these polarizations depends on the intrinsic parameters of the
binary, viz. the masses $m_1, m_2$ of the individual components
\footnote{Strictly, the masses that determine the shape of the GW signal
measured at the detector are the so-called ``detector-frame'' masses, which are
simply the masses multiplied by $(1 + z)$, where $z$ is the cosmological
redshift.}, and the spin vectors of these components $\spinvec_1,
\spinvec_2$. The amplitude of the polarizations depend on the inclination of
the binary with respect to the line of sight, $\iota$, luminosity distance
$d_L$, the phase $\phi_c$ and time $t_c$ at coalescence. In addition to
the above mentioned intrinsic parameters, we need two more parameters --
eccentricity $\eGeneric$ and mean anomaly $\lGeneric$~\footnote{The angle
variable that describes the position of the compact objects on the orbit can be
also described by other equivalent variables like eccentric anomaly or true
anomaly.~\cite{Clarke:2022fma}} -- at a given reference frequency to accurately
describe GWs from a binary in eccentric orbit. Also, the GW waveform shape is
more sensitive to the `chirpmass' $\Mc=\mchirp$ compared to individual masses
and often it is more practical to use $\Mc$ and the mass ratio $q=m_1/m_2$
instead of $m_1$ and $m_2$ for parameter estimation (PE) \cite{Cutler:1994ys}.

The response of the GW detector to each of the GW polarizations is given by the
antenna pattern functions, $F_{+}, F_{\times}$. These depend on the location of
the source in the sky, viz. the right ascension $\alpha$ and declination
$\delta$, as well as the polarization angle $\psi$. The GW strain measured by
the detector is a linear combination of the two polarizations, weighted by the
antenna pattern functions:
\begin{equation}
    h(t; \vec{\theta}, \vec{\lambda}) = F_{+}(\vec{\lambda}_F)h_{+}(t; \vec{\theta}, \vec{\lambda}_h) + F_{\times}(\vec{\lambda}_F)h_{\times}(t; \vec{\theta}, \vec{\lambda}_h)
\end{equation}
with $t$ as the detector-frame time, $\vec{\theta} = \lbrace{\Mc, q, \spinvec_1, \spinvec_2, \eGeneric, \lGeneric \rbrace}$, $\vec{\lambda}_F = \lbrace{\alpha, \delta, \psi \rbrace}$, $\vec{\lambda}_h = \lbrace{d_L, \iota, t_c, \phi_c \rbrace}$, and $\vec{\lambda} = \vec{\lambda}_F \cup \vec{\lambda}_h$.

Given two waveforms $h_{1,2}(t; \vec{\theta}, \vec{\lambda})$, their noise-weighted inner product is defined as:
\begin{equation}
    \langle h_{1}, h_{2} \rangle = 4\Re\int_{f_{\text{min}}}^{f_{\text{max}}}\frac{\tilde{h}^{\star}_1(f; \vec{\theta}, \vec{\lambda})\tilde{h}_2(f; \vec{\theta}, \vec{\lambda})}{S_n(f)}df
\end{equation}
where $f$ is the frequency, and $\tilde{h}_{1,2}$ are Fourier transforms of the
corresponding time domain waveforms, and $\star$ represents complex conjugation
and $\Re$ denotes the real part. $S_n(f)$ is the noise power spectral density (PSD)
of the detector. The lower and the upper cutoff frequencies, $f_{\text{min}}$
and $f_{\text{max}}$ are determined by the source properties and the sensitivity of the
detector.

Let us now consider a time-domain stretch of detector strain data:
\begin{equation}
    s(t) = n(t) + h(t; \vec{\theta}, \vec{\lambda})
\end{equation}
known to consist of noise $n(t)$, and GW signal $h(t; \vec{\theta}, \vec{\lambda})$
whose parameters $\vec{\theta}, \vec{\lambda}$ are to be estimated. This can be
achieved by sampling the seventeen-dimensional GW posterior $p(\vec{\theta},
\vec{\lambda} | s) \propto p(\vec{\theta}, \vec{\lambda})p(s | \vec{\theta},
\vec{\lambda})$, where $p(\vec{\theta}, \vec{\lambda})$ is the prior
distribution on the parameters, and the GW likelihood is given by
\cite{Cutler:1994ys}:
\begin{equation}
    p(s | \vec{\theta}, \vec{\lambda}) \propto \exp\left[-\frac{\langle s - h(\vec{\theta}, \vec{\lambda}), s - h(\vec{\theta}, \vec{\lambda}) \rangle}{2}\right]
\end{equation}
The sampling of this high-dimensional posterior is typically achieved using
large-scale Markov-chain Monte-Carlo methods, or nested-sampling methods.

\subsection{The inspiral-merger-ringdown consistency test}

The IMRCT can be thought of as a coarse-grained test of the frequency evolution
of a BBH waveform. It compares the low-frequency and high-frequency portions of
the waveform by comparing the final mass and spin of the merged binary
estimated from the inspiral and merger-ringdown regimes. Let $\massfInsp$ and
$\spinfInsp\equiv |S_f^\text{\insp}|/(M_f^\text{\insp})^2$ be the estimates of
the detector-frame final mass and dimensionless spin of the merged binary,
respectively, from the inspiral and $\massfPostInsp$, $\spinfPostInsp$ be the
same from the post-inspiral part. The IMRCT constructs a joint distribution on
the following difference parameters \cite{Ghosh:2016qgn, Ghosh:2017gfp}:
\begin{equation}
    \mdev \equiv 2\frac{\massfInsp - \massfPostInsp}{\massfInsp + \massfPostInsp},~ \sdev \equiv 2\frac{\spinfInsp - \spinfPostInsp}{\spinfInsp + \spinfPostInsp}
\end{equation}
The joint-distribution on these parameters should be consistent with zero if
the waveform model used to extract the signal in the data captures all the
physics associated with the production of the signal. On the other hand, a
sufficiently large deviation from GR is expected to shift this distribution to
an extent that it does not contain zero with high confidence ($\gtrsim 68\%$ or
$\gtrsim 90\%$). In this work, we demonstrate that even if both the waveform
model and the signal in the data are consistent with GR, an eccentric signal
recovered with a \qc model will also lead to a (spurious) violation
of the IMRCT.

The estimate of the final mass and final spin of the merged binary determined
with the intrinsic parameters $\vec{\theta}$ extracted from the inspiral
portion of the waveform is achieved via numerical-relativity-based fits
\cite{Hofmann:2016yih, Healy:2016lce, Jimenez-Forteza:2016oae}, as done in
\cite{Ghosh:2016qgn, Ghosh:2017gfp}. On the other hand, the shape of the
ringdown contained in the post-inspiral portion of the waveform depends on the
final mass and spin, which can therefore be extracted directly.

In this work, we restrict ourselves to non-precessing waveforms, i.e, waveforms
produced by binaries whose spin-vectors are aligned or antialigned with the
orbital angular momentum vector $\Lvec$ ($z$-axis). Moreover, we only consider
binaries with mass-ratios $q$ that are close to 1 ($q= m_1/m_2 \sim
1$). We therefore only need to use the leading order $(2, \pm 2)$ multipole,
since higher multipoles for the types of systems we consider will be
suppressed.

The choice of the transition frequency that separates the inspiral and
post-inspiral portions of the waveform is not unique. However, all
implementations of the IMRCT use some approximation of the frequency
corresponding to the innermost stable circular orbit (ISCO) of the binary. The
LVK estimates the ISCO using the medians of intrinsic parameters $\theta$
extracted from the full inspiral-merger-ringdown (IMR) waveform. Narayan et al \cite{Narayan:2023vhm} use
the medians of the final mass and spin of the merged binary estimated from the
full IMR waveform, similar to what was originally done in \cite{Ghosh:2016qgn,
Ghosh:2017gfp}. However, as pointed out in \cite{Ghosh:2016qgn, Ghosh:2017gfp,
Narayan:2023vhm}, the IMRCT is largely insensitive to moderate variations in
the choice of the transition frequency. In this work, we use the exact values
of $\vec{\theta}$ (i.e, the injected values), as well as the numerical
relativity fits \cite{Hofmann:2016yih, Healy:2016lce, Jimenez-Forteza:2016oae},
to determine a point-estimate of the final mass and spin, from which the ISCO
frequency is evaluated.

\subsection{A standardized definition of eccentricity}
\label{sec:A_standardized_definition_of_eccentricity}
In GR, there is no universally agreed-upon definition of
eccentricity. Consequently, different waveform models adopt custom internal
definitions of eccentricity, leading to model-dependent inferences about
eccentricity. To mitigate ambiguity in the inferred value of eccentricity from
gravitational wave data analysis, there is a growing effort to standardize the
definition of eccentricity. This standardization involves measuring
eccentricity from a gauge-independent quantity, such as the gravitational
waveform modes~\cite{Mora:2002gf, Ramos-Buades:2019uvh, Islam:2021mha,
Nagar:2021gss, Ramos-Buades:2021adz, Bonino:2022hkj, Ramos-Buades:2022lgf,
Shaikh:2023ypz}.

The waveform modes are obtained from the polarizations by decomposing the
complex combination $\hplus - i \hcross$ on a sphere into a sum of
spin-weighted spherical harmonic modes $\hlm$, so that the waveform along any
direction $(\iota, \varphi_0)$ in the binary's source frame is given by
\begin{equation}
\label{eq:h}
\h(t, \iota, \varphi_0) =
\sum_{\ell=2}^{\ell=\infty}\sum_{m=-\ell}^{m=\ell} \hlm(t) ~
{}_{-2}Y_{\ell m}(\iota,\, \varphi_0),
\end{equation}
where $\iota$ and $\varphi_0$ are the polar and azimuthal angles on the sky in
the source frame, and ${}_{-2}Y_{\ell m}$ are the spin $=-2$ weighted spherical
harmonics.

In this work, we employ the \texttt{Python} implementation \package
of the standardized definition of eccentricity from \cite{Shaikh:2023ypz},
which defines the eccentricity $\egw$ using the frequency of the \dm mode
$\htwotwo$. Initially, it computes $\etwotwo$ as described in
\cite{Ramos-Buades:2019uvh, Islam:2021mha, Bonino:2022hkj,Shaikh:2023ypz}.
\begin{equation}
  \label{eq:e_omegatwotwo}
  \etwotwo(t) = \frac{\sqrt{\omegaP(t)} -
    \sqrt{\omegaA(t)}}{\sqrt{\omegaP(t)} + \sqrt{\omegaA(t)}},
\end{equation}
Here, $\omegaP(t)$ and $\omegaA(t)$ denote the interpolants through the
frequency of the \dm mode at the pericenter (point of closest approach) and the
apocenter (point of farthest approach), respectively. The frequency
$\omegatwotwo$ of the \dm mode is derived from $\htwotwo$ as
\begin{align}
\htwotwo(t) = \Atwotwo(t)\,e^{- i \phitwotwo(t)}, \\
\omegatwotwo(t) = \frac{\dd \phitwotwo(t)}{\dd t}\,.
\label{eq:omega_22}
\end{align}
After computing $\etwotwo$ from $\htwotwo$, $\egw$ is determined through the
following transformation to ensure that $\egw$ correctly exhibits the Newtonian
limit~\cite{Ramos-Buades:2022lgf}.
\begin{equation}
\label{eq:eGW}
  \egw = \cos(\Psi/3) - \sqrt{3} \, \sin(\Psi/3),
\end{equation}
where
\begin{equation}
  \label{eq:psi}
  \Psi = \arctan\left(\frac{1 - \etwotwo^2}{2\,\etwotwo}\right).
\end{equation}

Since eccentricity decays via gravitational wave radiation, we must specify a
reference frequency where the value of $\egw$ is measured. Because eccentricity
induces modulations in the instantaneous frequency $\omegatwotwo$, we obtain a
monotonic orbit-averaged $\avgOmega$ by taking the orbital average of the
instantaneous $\omegatwotwo$ (see Section II.E.2 of \cite{Shaikh:2023ypz} for
more details) Between any two consecutive extrema (local maxima (pericenters,
$\tP$) or minima (apocenters, $\tA$) in the instantaneous $\omegatwotwo$)
$\tX_i$ and $\tX_{i+1}$ one can define
\begin{align}
  \avgOmega_{i}^{\text{X}} & = \frac{1}{\tX_{i+1} -
                  \tX_{i}} \, \int_{\tX_{i}}^{\tX_{i+1}}\omegatwotwo(t)\,\dd t
                  \nonumber \\
\label{eq:mean_motion}
& = \frac{\phitwotwo(\tX_{i+1}) - \phitwotwo(\tX_{i})}{\tX_{i+1} - \tX_{i}},
\end{align}
and associate $\avgOmega_i^\text{X}$ with the midpoint between $\tX_{i}$
and $\tX_{i+1}$:
\begin{equation}
  \avgT^{X}_{i} = \frac{1}{2}\left(\tX_i + \tX_{i+1}\right).
\end{equation}
Applying this procedure to all consecutive pairs of pericenter times, one can
obtain the set $\big\{\,\left(\avgTP_i, \avgOmega_i^\text{p}\right)\,\big\}$
and similarly, the set
$\left\{\,\big(\avgTA_i,\avgOmega_i^{\text{a}}\big)\,\right\}$ for apocenter
times. Finally, taking the union of these two datasets, one can build an
interpolant in time to obtain $\avgOmega(t)$ and use $\avgfref (\tref) \equiv
\avgOmega (\tref) / (2\pi)$ for defining the reference frequency $\avgfref$
corresponding to a reference time $\tref$.

Presenting our results with a standardized definition facilitates
one-to-one comparisons with similar studies in the past and enables future
studies to do the same with the results in this work.

\subsection{Parameter Estimation setup}
\label{sec:Parameter_estimation_setup}
We obtain our results by performing Bayesian parameter inference runs using
state-of-the-art GW parameter inference framework built in \bilby with a
dynamical nested sampler \texttt{dynesty}~\cite{Speagle:2019ivv}. For the
injection and recovery, we use the publicly available eccentric waveform model
\TEOB~\cite{Nagar:2018zoe,Nagar:2020xsk,Chiaramello:2020ehz,Albanesi:2021rby,Nagar:2021gss,Nagar:2021xnh,Nagar:2024dzj}~\footnote{We
use the commit 0f19532 of the eccentric branch of \TEOBGeneric available at
\url{https:// bitbucket.org/eob_ihes/teobresums/}}. In the following, we will
denote the eccentricity that the waveform model takes as an input by $\eEOB$
and the frequency where $\eEOB$ is defined by $\fref$. Additionally, we provide
\{$\avgfref, \egw$\} for each injection to facilitate better comparisons across
waveform models and systematics studies.~\footnote{Note that we sample over the
model eccentricity $\eEOB$ and not $\egw$.}  For a given model eccentricity
$\eEOB$ at a model reference frequency $\fref$, the standardized eccentricity
$\egw$ at $\avgfref$ is obtained by first generating the eccentric waveform
modes for the given pair $\{\fref, \eEOB \}$ and then following the steps
discussed in Sec.~\ref{sec:A_standardized_definition_of_eccentricity}.

Ideally, to accurately describe an eccentric orbit, the waveform model should
also include mean anomaly as a free
parameter~\cite{Islam:2021mha,Clarke:2022fma,Shaikh:2023ypz}. In this work, we
use the version of \TEOB that was used in Ref.~\cite{Bonino:2022hkj} by the
waveform developers to showcase the accuracy and robustness of their
model. This version of \TEOB does not allow mean anomaly as a free
parameter. Although the most recent version of \TEOB, released after the
completion of this work, allows mean anomaly as a free parameter, this version
is yet to be tested (at the time of writing) in a parameter inference study by
sampling over mean anomaly and is currently under review. Eccentric waveform
models \texttt{Ecc1dSur}~\cite{Islam:2021mha} and
\SEOB~\cite{Ramos-Buades:2021adz,Ramos-Buades:2023yhy} include mean anomaly as
a free parameter but these are not publicly available. Publicly available
eccentric waveform models \SEOBNRE~\cite{Cao:2017ndf,Liu:2019jpg} and
\EccentricTD~\cite{Tanay:2016zog} do not include mean anomaly as a free
parameter but instead start at the pericenter or apocenter.\footnote{In some
classes of GW sources, eccentricity and mean anomaly may not be sufficient to
describe the orbit. For example, in Ref.~\cite{Carullo:2023kvj}, the
Arnowitt-Deser-Misner energy and angular momentum are used to describe a
generic planar orbit. See also Ref.~\cite{Gamba:2021gap} which use the same
parameterization to study the possibility of GW190521 resulting from a
dynamical capture of two nonspinning black holes.}

Because the version of \TEOB used in the work does not allow sampling over mean
anomaly, Ref.~\cite{Bonino:2022hkj} sampled over the reference frequency, i.e.,
the PE was performed by sampling over eccentricity and reference frequency with
fixed mean anomaly. The motivation behind such an approach is that by varying
the reference frequency with fixed mean anomaly one can expect to explore the
same parameter space as in the case of varying mean anomaly at a fixed reference
frequency. For example, in a recent work in Ref.~\cite{Bonino:2024xrv}, it was
shown that one can map an eccentric Numerical Relativity (NR) waveform to an eccentric effective one
body waveform using just the initial eccentricity and reference frequency at
the first apocenter (that is, at a fixed mean anomaly).  Although sampling
over reference frequency may reduce bias in PE compared to a fixed reference
frequency when mean anomaly is fixed, no significant bias was found in the
recovered parameters when sampling exclusively over eccentricity with both mean
anomaly and reference frequency fixed (see Fig.~3. in
Ref.~\cite{Bonino:2022hkj}).

  Contrary to Ref.~\cite{Bonino:2022hkj,Bonino:2024xrv} which suggest that
using (eccentricity, mean anomaly) at fixed reference frequency and using
(eccentricity, reference frequency) at fixed mean anomaly is equivalent, it
has also been argued that sampling over eccentricity and reference frequency
with fixed mean anomaly may not be equivalent to sampling over eccentricity and
mean anomaly with a fixed reference frequency (see the introduction in
Ref.~\cite{Gupte:2024jfe}) since these two methods may not explore the same
parameter space. In addition, varying reference frequency changes the reference
point where binary properties are measured.

Given the fact that no significant bias was reported by the waveform
developers in Ref.~\cite{Bonino:2022hkj} when sampling only eccentricity with
fixed reference frequency and that this version does not allow varying mean
anomaly, we restrict our PE to sample only over eccentricity. However, future
works that will use the recent version of \TEOB once it is reviewed, should
sample over mean anomaly to reduce any potential bias due to neglecting the mean
anomaly.

To obtain the 2D posteriors on $\mdevinline$ and $\sdevinline$ for the IMRCT,
we use the \texttt{summarytgr} command line tool from
\texttt{PESummary}~\cite{Hoy:2020vys}. It provides the posteriors on
$\mdevinline$ and $\sdevinline$ in a post-processing step making use of mass
and spin posterior samples along with the NR-fits for the final mass and the
final spin parameters. The uninformative priors used for the mass and spin
parameters force non-trivial undesirable priors on $\mdevinline$
and $\sdevinline$ obtained in the post-processing. To revoke this
effect and to impose agnostic priors on these inference parameters used to
probe GR-violation, we reweight\footnote{Reweighting of posterior
refers to removing the effect of any non-trivial undesirable prior on the
posterior (see Chapter 5 in Ref.~\cite{gelman2013bayesian} for
details). This is done when there is no prior information about the parameter
in consideration or the prior information about the parameter is unphysical and
one wants to be as objective as possible.} the 2D posterior to a flat prior on
the deviation parameters, as is routinely done by the LVK
\cite{LIGOScientific:2020tif,LIGOScientific:2021sio}.

Given the posterior samples obtained in the PE from the inspiral and
post-inspiral part, \texttt{summarytgr} uses an NR informed fit to obtain the
final mass and spin posteriors. This fit is based on \qc NR simulations only
and does not take into account any eccentric corrections which has been shown
to be negligible at the quasi-Newtonian level with small eccentricity
approximation~\cite{Bhat:2022amc}. Currently, we lack a sufficient number of
NR-simulations of merging eccentric binaries required for obtaining the
eccentricity corrected NR-fit for final mass and final spin. We measure
\{$\avgfref, \egw$\} for a given \{$\fref, \eEOB$\} using
\package~\cite{Shaikh:2023ypz}.

\section{Results}
\label{sec:results}
The main result of this work can be divided into two parts. The first part,
Sec.~\ref{sec:Effects_of_neglecting_eccentricity_in_templates_on_IMRCT},
discusses the effect of neglecting eccentricity in the waveform template on the
IMRCT. The second part, Sec.~\ref{sec:IMRCT_on_eccentric_signal}, demonstrates
the robustness of the IMRCT performed on GW data containing an eccentric signal
using a waveform template that includes physics required to describe GWs coming
from binaries on an eccentric orbit. We list the injection values
and priors in \ref{tab:aligned_injection} and summarize the results of the study
of bias in IMRCT due to neglecting eccentricity in \ref{tab:bias_summary}.

\subsection{Effects of neglecting eccentricity in templates on IMRCT}
\label{sec:Effects_of_neglecting_eccentricity_in_templates_on_IMRCT}
\begin{figure*}
  \centering
  \includegraphics{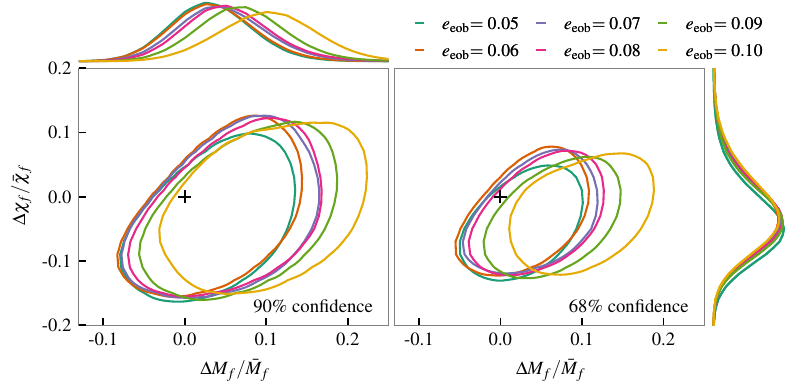}\\
  \includegraphics{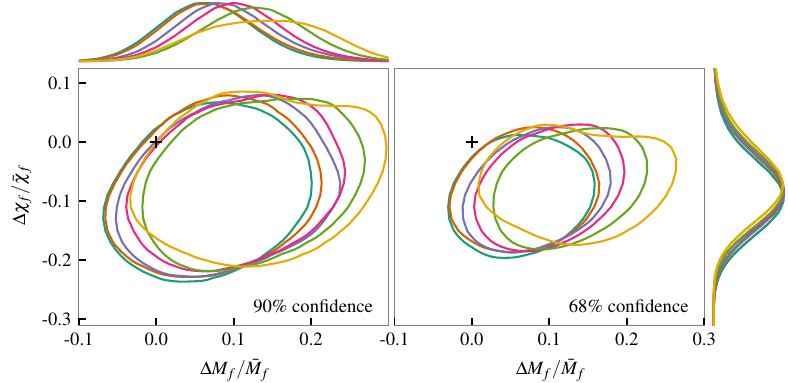}
  \caption{Bias in IMRCT due to neglecting eccentricity in recovery
template. The top and the bottom row represent the IMRCT result for the aligned
spin configurations with dimensionless component spins $\spinpair = (0.4, 0.3)$
and $\spinpair=(0.2, 0.1)$, respectively. The contours represent the joint
probability distribution of the final mass deviation $\mdevinline$ and final
spin deviation $\sdevinline$ at 90\% and 68\% confidence on the left and the
right panels, respectively. For $\spinpair=(0.4, 0.3)$, GR value is excluded at 68\% confidence at $\eEOB = 0.09$. For $\spinpair=(0.2, 0.1)$, GR value is excluded at 68\% and 90\% confidence for $\eEOB=0.05$ and $0.07$, respectively.}
\label{fig:bias_due_to_quasicircular}
\end{figure*}

\subsubsection{Aligned spin case}
\label{sec:bias_aligned_spin_case}
First, we focus on studying the cases where the spins of the binary components are
aligned with the orbital angular momentum of the binary. It is convenient to define the
tilts $\tilt_{i}$ as the angle between of the component spin $\spinvec_{i}$ and the orbital angular momentum $\Lvec$

\begin{equation}
  \label{eq:tilt}
  \tilt_{i} = \cos^{-1}\left(\frac{\Lvec.\spinvec_{i}}{|\Lvec||\spinvec_{i}|}\right)\qquad i=\{1,2\}
\end{equation}
Since \TEOB can describe only non-precessing systems, $\tilt_{i}$ can take only
two values $0$ (aligned) or $\pi$ (antialigned). In this section, we consider
binary systems with $\tilt_{1,2} = 0$. Since the waveform model only allows tilts of either 0 or $\pi$, and does not allow sampling over arbitrary spin orientations, we choose the prior on the tilts to be a Dirac delta function peaked at the injection value.

We choose the component masses to be similar to that of the gravitational wave
event GW150914. For this study we choose a total mass of system that lies
within the `golden mass' range $M \in [50-100] M_{\odot}$ which ensures high
enough SNRs both in the inspiral and post-inspiral parts of the
signal~\cite{Ghosh:2016qgn}.  The injection parameters and the priors are noted
in Table~\ref{tab:aligned_injection}. The injection eccentricities $\eEOB \in$
\{0.05, 0.06, 0.07, 0.08, 0.09, 0.1\} are defined at a reference frequency of
$\fref=$ 20 Hz. These injected waveforms have $\egw$ values \{0.040, 0.048,
0.055, 0.063, 0.071, 0.079\}~\footnote{We used the \mResAmp method within
\package to measure these eccentricities.} measured at an orbit averaged
frequency of $\avgfref=$ 25 Hz. To understand the effect of spin magnitudes on IMRCT systematics, we consider two different choices of dimensionless component
spins, one low $\spinpair = (0.2, 0.1)$ and one moderate $\spinpair = (0.4,
0.3)$, for the set of injected eccentricities. All the injections considered in
this study have Hanford-Livingston-Virgo network (at O4 design sensitivity) signal to noise ratio (SNR) in the range $27-52$.

While recovering the injection, we use a Dirac
delta prior on eccentricity peaked at $\eEOB=10^{-4}$ making the parameter
estimation procedure equivalent to one using a \qc template.~\footnote{For
$\eEOB \leq 10^{-4}$, the corresponding $\egw$ is $10^{-6}$ implying the
waveforms to be essentially \qc. See fig. 10 in ~\cite{Shaikh:2023ypz}}

\begin{table}
\begin{ruledtabular}
\begin{tabular}{c|c|c}
  Parameter & Injected value & Prior \\\hline
  $\mathcal{M}_c (M_\odot)$ & 26.36 & $\mathcal{U}(12, 45)$\\
  $q$ & 0.8 & $\mathcal{U}(0.25, 1)$\\
  $\spin_{1}$ & \{0.2, 0.4\} &$\mathcal{U}(0, 0.99)$\\
  $\spin_{2}$ & \{0.1, 0.3\} &$\mathcal{U}(0, 0.99)$\\
  $\tilt_1$ (rad) & 0 & 0 \\
  $\tilt_2$ (rad) & 0 & 0 \\
  $d_L$(MPc) & 410 &$\mathcal{U}(50, 2000)$\\
  $\iota$ & $0.78$ &$\mathcal{\sin}(0, \pi)$\\
  $\alpha$(rad) & 4.89 &$\mathcal{U}(0, 2\pi)$\\
  $\delta$ & -0.218 &$\mathcal{\cos}(-\pi/2, \pi/2)$\\
  $\psi$(rad) & 0.54 &$\mathcal{U}(0, \pi)$\\
  $t_c$(s) & $1126259462$ ($\equiv t_0$) &$\mathcal{U}(t_0-2, t_0+2)$\\
  $\phi_c$(rad) & 4.205 &$\mathcal{U}(0, 2\pi)$\\
  $\eEOB$ & \{0.05, 0.06, 0.07, 0.08, 0.09, 0.10\} &$10^{-4}$\\
  $\fref$(Hz) & 20 & 20
\end{tabular}
\end{ruledtabular}
\caption{\label{tab:aligned_injection} Injection values of the binary
parameters and corresponding priors used for PE with
\bilby. $\mathcal{U}(a, b)$ stands for uniform prior in the range $(a,
b)$. $\mathcal{\sin}(0, \pi)$ and $\mathcal{\cos}(-\pi/2, \pi/2)$ stand for the
Sine and Cosine priors, respectively on the ranges mentioned. Single values
represent Dirac delta priors peaked at those values. Different values in the
the curly braces represent different injections. The table, therefore,
represents 12 injections -- 6 different eccentricities each with 2
different spin configurations.}
\end{table}

The results of these IMRCTs are shown in
Fig.~\ref{fig:bias_due_to_quasicircular}. The top row shows the results for
$\spinpair = (0.4, 0.3)$ and the bottom row shows the same for $\spinpair =
(0.2, 0.1)$. In each of these rows, the left panel shows the iso-probability
contours of the 2D probability distribution on the final mass deviation
$\mdevinline$ (along $x$-axis) and the final spin deviation $\sdevinline$
(along $y$-axis) plane at 90\% confidence. The right panel shows the same at
68\% confidence. Different contours in each panel correspond to the injected
eccentricities $\eEOB\in$ \{0.05, 0.06, 0.07, 0.08, 0.09, 0.1\} as noted on the
top of the right panel. Fig.~\ref{fig:bias_due_to_quasicircular} also shows the
marginalized 1D posterior of $\mdevinline$ on top of the left panel and that of
$\sdevinline$ on the right of the right panel.

It can be noted that in the case of moderate spin $\spinpair = (0.4, 0.3)$, for
smaller eccentricities ($\eEOB < 0.08$), the contours at 90\% and 68\% contain
the point (0, 0) denoted by $+$ which stands for zero-deviation from GR. We
will denote this point of zero-deviation from GR as the `GR value'. However,
for larger eccentricities ($\eEOB \gtrsim 0.08$), the contours are shifted towards
positive deviation in $\mdevinline$ and the GR value is excluded at 68\%
confidence for $\eEOB \geq 0.09$. For smaller spins $\spinpair = (0.2, 1)$,
the bias is more severe. For the low spin case (bottom row in
Fig.~\ref{fig:bias_due_to_quasicircular}), the GR value is excluded at 68\%
confidence for $\eEOB = 0.05$ and at 90\% confidence for $\eEOB = 0.07$.

\begin{figure}
  \centering
  \includegraphics[width=\columnwidth]{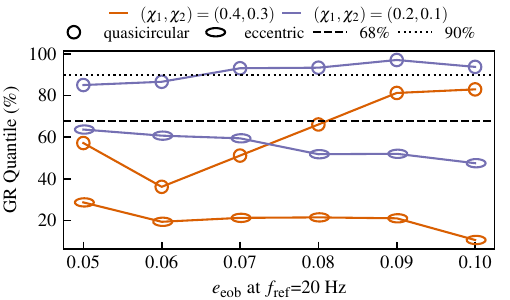}
  \caption{GR Quantile $\QGR$ vs injected eccentricity for different component
spins for the aligned spin systems. Small $\QGR$ indicates better agreement
with GR. Circles denote the \qc recovery and the ellipses denote the
eccentric recovery. The low spin $\spinpair = (0.2, 0.1)$ and moderate spin
$\spinpair = (0.4, 0.3)$ case is denoted by violet and orange,
respectively. When recovered with \qc templates, low spin cases show
more bias compared to moderate spin case.}
  \label{fig:gr_quantile_aligned_spin}
\end{figure}

One can also construct a statistical measure of the deviation from GR using the
GR quantile $\QGR$ which is the fraction of the posterior samples contained
within the iso-probability contour that passes through the GR value
(no-deviation point $(0, 0)$). A smaller $\QGR$ implies a better agreement with
GR since the GR value is contained within a smaller iso-probability contour.
In this work, we compute the GR quantile in a slightly different way. Instead
of using the samples directly, we use an interpolated histogram of the 2D
probability distribution. In particular, we use the implementation in
\texttt{summarytgr} within \texttt{PESummary}~\cite{Hoy:2020vys}.

In
Fig.~\ref{fig:gr_quantile_aligned_spin}, we plot the $\QGR$ for different
component spin configurations as a function of the injection eccentricities. It shows
the overall trend of $\QGR$ increasing with $\eEOB$. While for moderate spins
$\spinpair = (0.4, 0.3)$, $\QGR$ becomes $\approx 68\%$ at $\eEOB = 0.08$, for
low spins $\spinpair = (0.2, 0.1)$, $\QGR$ becomes $ > 68\%$ at $\eEOB = 0.05$
and $\gtrsim 90\%$ at $\eEOB = 0.07$.

\subsubsection{Antialigned spin case}
\label{sec:bias_anti_aligned_spin_case}
\begin{figure*}
  \centering
  \includegraphics{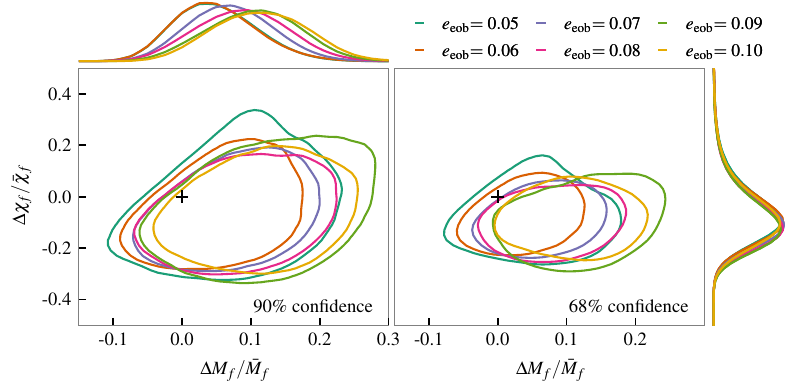}\\
  \includegraphics{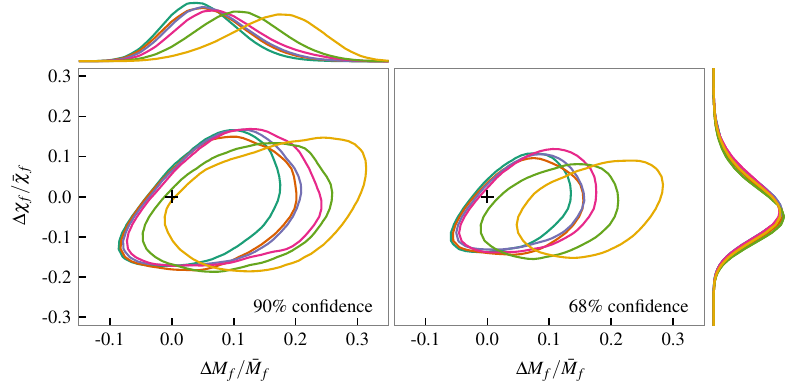}\\
  \includegraphics{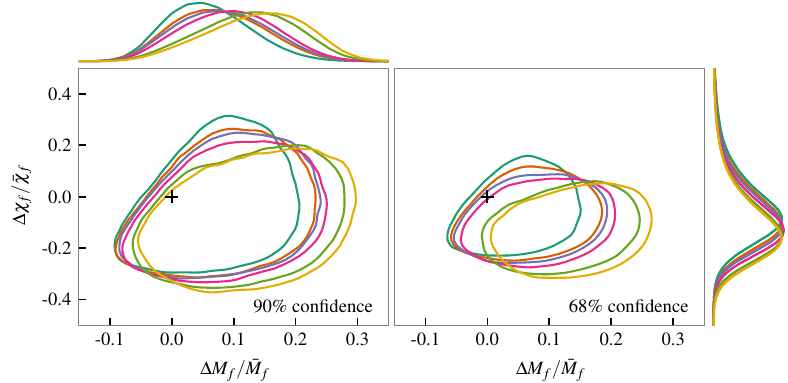}
  \caption{IMRCT with \qc recovery template for injections with antialigned
spin configurations.The panels from the top to bottom correspond to primary
antialigned, secondary antialigned and both antialigned spin configurations,
respectively. For all the three cases, the dimensionless component spin magnitudes
are $(\spin_1, \spin_2)=(0.4, 0.3)$. Due to neglecting eccentricity in the
recovery template, the GR value is excluded at 68\% confidence at $\eEOB=0.07,
0.09$ and $0.08$ for primary antialigned, secondary antialigned and both
antialigned cases, respectively. The GR value is also excluded at 90\%
confidence for secondary antialigned at $\eEOB=0.1$.}
  \label{fig:imrct_with_quasicircular_antialigned}
\end{figure*}

In this section, we study how the bias in the IMRCT due to neglecting
eccentricity in the recovery waveform model depends on the orientation of the
spin with respect to the angular momentum of the system. Let the pair
$(\tilt_1, \tilt_2)$ denote the tilts of the spin $\spinvec_1$ of the primary
and the spin $\spinvec_2$ of the secondary, respectively, with respect to the
orbital angular momentum $\Lvec$ of the system. We consider the following three cases
\begin{enumerate}
\item $(\pi, 0)$: $\spinvec_1$ is antialigned and $\spinvec_2$ is aligned.
\item $(0, \pi)$: $\spinvec_1$ is aligned and $\spinvec_2$ is antialigned.
\item $(\pi, \pi)$: Both $\spinvec_1$ and $\spinvec_2$ are antialigned.
\end{enumerate}
We consider only the moderate spin case where the dimensionless spins are (0.4,
0.3).  The other parameters remain the same as in Table
\ref{tab:aligned_injection}. We use a Dirac delta prior peaked at the injected
values for the tilts. The number of injections for each of these tilts is 6 for
six different eccentricities making a total of 18 injections for these three
tilt configurations.

The rows from the top to bottom in
Fig.~\ref{fig:imrct_with_quasicircular_antialigned} show the IMRCT results for
the cases of primary antialigned, secondary antialigned and both antialigned,
respectively. In each of these figures, the left panel shows the
iso-probability contours at 90\% confidence and the right panel shows the same
at 68\% confidence in the $\mdevinline-\sdevinline$ plane. The general trend is
similar to that of aligned spin cases in
Sec.~\ref{sec:bias_aligned_spin_case}.

\begin{table*}
\begin{ruledtabular}
\begin{tabular}{c|c|c|c|c|c}
 Dimensionless component spins & Tilts & \multicolumn{2}{c|}{Threshold eccentricity for GR violation at $90\%$} & \multicolumn{2}{c}{Threshold eccentricity for GR violation $68\%$} \\\hline
  $(\spin_1, \spin_2)$  & $(\tilt_1, \tilt_2)$ & $\eEOB$ at $\fref=20$Hz & $\egw$ at $\avgfref=25$Hz & $\eEOB$ at $\fref=20$Hz & $\egw$ at $\avgfref=25$Hz \\\hline
  (0.4, 0.3) & (0, 0) &  &  & 0.09 & 0.071 \\ \hline
  (0.2, 0.1) & (0, 0) & 0.07 & 0.055 & 0.05 & 0.040 \\ \hline
  (0.4, 0.3) & $(\pi, 0)$ & & & 0.07 & 0.056 \\ \hline
  (0.4, 0.3) & $(0, \pi)$ & 0.10 & 0.079 & 0.09 & 0.071 \\ \hline
  (0.4, 0.3) & $(\pi, \pi)$ &  & & 0.08 & 0.064 \\
\end{tabular}
\end{ruledtabular}
\caption{Summary of the study of bias in IMRCT due to neglecting
eccentricity. It provides the threshold injection eccentricity ($\eEOB$ at 20
Hz) at or above which the IMRCT result is biased at 68\% and 90\%
confidence. It also provides the eccentricity $\egw$ in terms of standardized
definition~\cite{Shaikh:2023ypz} at an orbital averaged frequency $\avgfref=25$
Hz corresponding to each of these threshold injections. As an example, for an
aligned low spin $\spinpair = (0.2, 1.0)$ system, GR is excluded at 68\%
confidence for $\eEOB=0.05$ ($\egw = 0.04$) and GR is excluded at 90\%
confidence for $\eEOB=0.07$ ($\egw=0.055$).}
\label{tab:bias_summary} 
\end{table*}

\subsection{IMRCT with eccentric signal}
\label{sec:IMRCT_on_eccentric_signal}
In the previous section, we showed that IMRCT shows spurious deviation from GR
when the parameters from an eccentric gravitational wave signal is inferred
using a \qc waveform model (using a delta function prior on
eccentricity fixed at $\eEOB=10^{-4}$). In this section, we show that such
spurious deviations go away when eccentric waveform model is used instead of
\qc model by allowing the sampler to sample from a uniform prior on
eccentricity as described in Tab.~\ref{tab:eccentric_recovery}.
\begin{table}
\begin{ruledtabular}
\begin{tabular}{c|c|c}
  Parameter & Injected value & Prior \\\hline
  $\eEOB$ & \{0.05, 0.06, 0.07, 0.08, 0.09, 0.10\} &$\mathcal{U}(8 \times 10^{-3}, 0.15)$\\
  $\fref$(Hz) & 20 & 20
\end{tabular}
\end{ruledtabular}
\caption{Injection values of the eccentricities and priors for IMRCTs with
eccentric recovery.}
\label{tab:eccentric_recovery}
\end{table}

\subsubsection{Aligned spin case}
\label{sec:eccentric-Aligned_spin_case}
\begin{figure*}[!]
  \centering
  \includegraphics{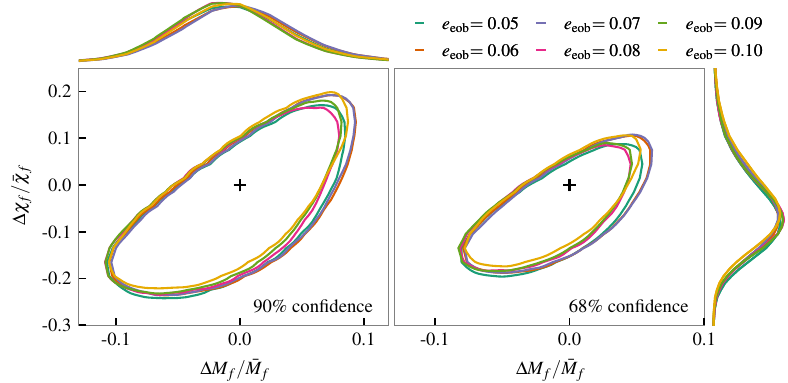}\\
  \includegraphics{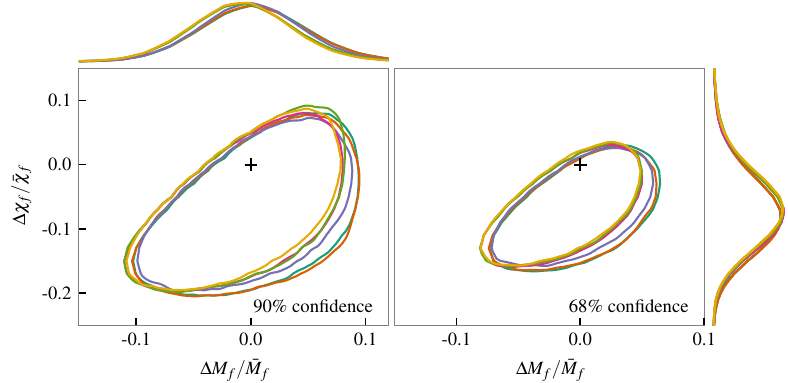}
  \caption{IMRCT with eccentric recovery template for aligned spin
injections. The top row corresponds to $(\spin_1, \spin_2)=(0.4, 0.3)$ and
the bottom row corresponds to $(\spin_1, \spin_2)=(0.2, 0.1)$. Same as
Fig.~\ref{fig:bias_due_to_quasicircular}, but now the injections are recovered
with eccentric templates instead of \qc templates. GR value is included within
the iso-probability contours at both the 68\% and 90\% confidence intervals for
all injections for these two spin configurations.}
  \label{fig:imrct_with_eccentric_template}
\end{figure*}
First, we consider the case of aligned spin where both the primary and the
secondary spin are aligned with the angular momentum of the system, i.e.,
$\tilt_{1,2} = 0$. In Fig.~\ref{fig:imrct_with_eccentric_template}, we show the
results of IMRCTs with a set of eccentric aligned spin signals. The top row
shows the result for injections with moderate component spins $(\spin_1,
\spin_2)=(0.4, 0.3)$ and the bottom row shows the result for low component
spins $(\spin_1, \spin_2)=(0.2, 0.1)$. The set of injected eccentricities and
the prior on the eccentricity for both of these spin configurations are
provided in Tab.~\ref{tab:eccentric_recovery}. The other parameters remain the
same as in Sec.~\ref{sec:bias_aligned_spin_case}.

In each row of Fig.~\ref{fig:imrct_with_eccentric_template}, the left (right)
panel shows the iso-probability contours of the 2D probability distribution in
the $\mdevinline-\sdevinline$ plane at 90\% (68\%) confidence. The 1D
posterior on $\mdevinline$ is shown on top of left panel and the 1D posterior on
$\sdevinline$ is shown on the right of the right plot. In both of these spin
configurations, for all the injected eccentricities, we find IMRCTs to be
unbiased both at 68\% and 90\% confidence.

For these two aligned spin configurations, the GR quantile $\QGR$, when
recovered with eccentric waveform template, is shown using ellipses in
Fig.~\ref{fig:gr_quantile_aligned_spin}. $\QGR$ is plotted as a function of the
injection eccentricities. It can be seen that for all the injected
eccentricities, the $\QGR$ remains $< 68\%$ for low spins $(\spin_1,
\spin_2)=(0.2, 0.1)$ and $< 30\%$ for moderate spins $(\spin_1, \spin_2)=(0.4,
0.3)$ implying an agreement with GR.

\subsubsection{Antialigned spin case}
\label{sec:eccentric-antialigned_spin_case}
\begin{figure*}[!]
  \centering
  \includegraphics{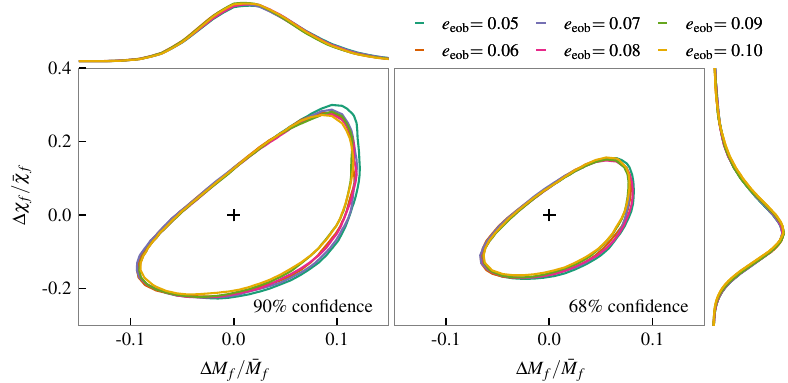}\\
  \includegraphics{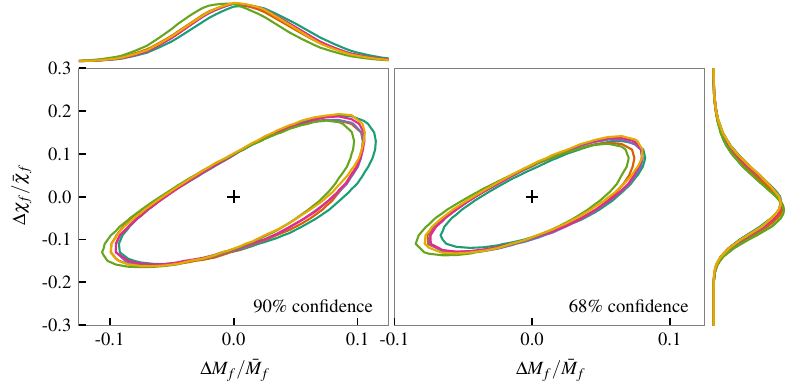}\\
  \includegraphics{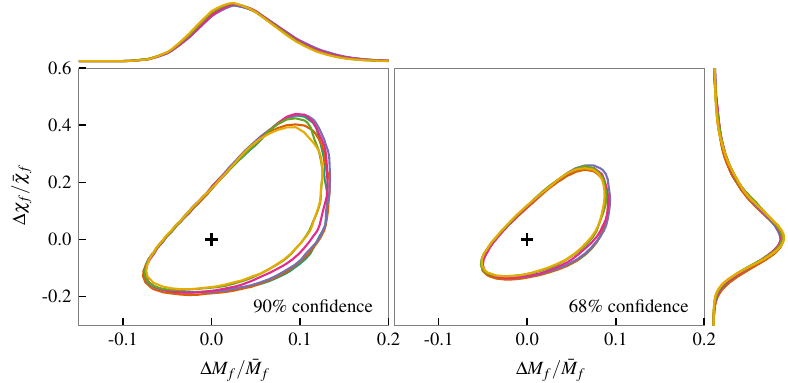}
  \caption{IMRCT with eccentric recovery template for injections with
antialigned spin configurations. The rows from the top to bottom correspond to
primary antialigned, secondary antialigned and both antialigned spin
configurations, respectively. For all the three cases, the dimensionless component
spin magnitudes are $(\spin_1, \spin_2)=(0.4, 0.3)$. Same as
Fig.~\ref{fig:imrct_with_quasicircular_antialigned} but now the injections are
recovered with eccentric templates instead of \qc templates. Note that the GR
value is included at both 68\% and 90\% confidence for all the injections for these
three configurations.}
  \label{fig:imrct_with_eccentric_antialigned}
\end{figure*}
We now turn our focus to the potential effects of the orientation of the
component spins on IMRCTs with eccentric signal when recovered with eccentric
waveform models. We consider the same injections as in
Sec.~\ref{sec:bias_anti_aligned_spin_case}; we consider three possible
configurations where the pair of tilts may take the values $(\pi, 0), (0, \pi)$
or $(\pi, \pi)$ with the component dimensionless spins $(\spin_1,
\spin_2)=(0.4, 0.3)$. In Sec.~\ref{sec:bias_anti_aligned_spin_case}, these
injections were recovered using a \qc template which resulted in spurious bias
in the IMRCT result. Now we recover these injections with eccentric template
using a prior on eccentricity as shown in Tab.~\ref{tab:eccentric_recovery}.

The rows from the top to bottom in
Fig.~\ref{fig:imrct_with_eccentric_antialigned} shows the IMRCT contours for
primary antialigned, secondary antialigned and both antialigned configurations,
respectively. For a given row, the left (right) panel shows the iso-probability
contours of the 2D probability distribution in the $\mdevinline-\sdevinline$
plane at 90\% (68\%) confidence. The 1D posterior on $\mdevinline$ is shown on
top of the left panel and the 1D posterior on $\sdevinline$ is shown on the
right of the right panel. Different contours in each panel correspond to
different injected eccentricity as denoted on top of the right panel of each
row. Unlike the results in Fig.~\ref{fig:imrct_with_quasicircular_antialigned},
all the contours in Fig.~\ref{fig:imrct_with_eccentric_antialigned} include the
GR value.

We also plot the $\QGR$ (denoted by ellipses) values of the IMRCTs
corresponding to these injections in
Fig.~\ref{fig:gr_quantile_antialigned_spin}. Note that the $\QGR$ values for
these cases of recovery with eccentric templates are very small compared to the
\qc recovery (denoted by circles). The $\QGR$ for eccentric recovery is $<
68\%$ for all considered eccentricities in these three antialigned spin
configurations. While for the primary antialigned cases, the $\QGR$ remains
$\lesssim 10\%$, for the secondary antialigned and both antialigned cases, the
GR quantile is even smaller, $\QGR \lesssim 5\%$ implying excellent agreement
with GR.

\begin{figure}
  \centering
  \includegraphics[width=\columnwidth]{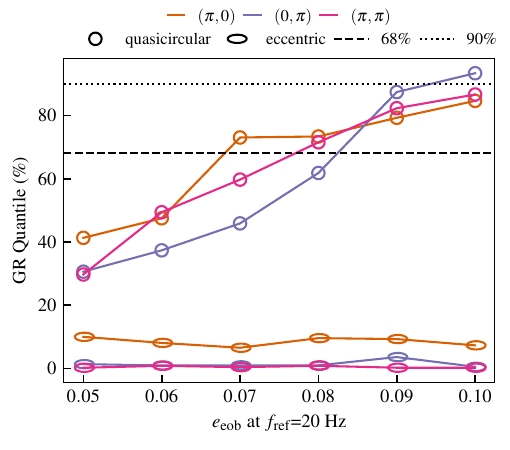}
  \caption{GR Quantile $\QGR$ vs injected eccentricity for different component
spin alignments for $\spinpair = (0.4, 0.3)$. Small $\QGR$ indicates better
agreement with GR. Circles denote the \qc recovery and the ellipses denote the
eccentric recovery. When recovered with \qc template, the $\QGR$ tend to
increase with increasing eccentricity implying increasing bias in IMRCT due to
neglecting eccentricity in the waveform model. On the other hand, for eccentric
recovery, the $\QGR$ remains within $68\%$ for all eccentricities. For all
cases of eccentric recovery, the $\QGR$ is $\lesssim 10\%$. In the secondary
antialigned and both antialigned cases, the $\QGR$ becomes even smaller ($<
5\%$) implying excellent agreement with GR.}
  \label{fig:gr_quantile_antialigned_spin}
\end{figure}

\section{Summary and Outlook}
\label{sec:conclusion}
The IMRCT \cite{Ghosh:2016qgn, Ghosh:2017gfp} is one among several tests of GR
with GWs used by the LVK. While model-agnostic in the sense that it is not
based on any particular alternative theory of gravity, it does nevertheless
crucially rely on the CBC template waveform being an accurate representation of
the GW signal. Thus, intuitively, one would expect spurious violations of this
test when physical and environmental effects are not accounted for in the
waveform model.

To demonstrate this, in this work, we focus our attention on GWs from eccentric
compact binaries. We inject, in zero-noise assuming an O4-like noise PSD,
synthetic GW150914-like CBC signals but with a range of residual eccentricities
at a reference frequency of $f_{\mathrm{ref}} = 20$ Hz. The spins are always
assumed to lie along the orbital angular momentum ($z-$) axis, although we
consider two different sets of spin magnitudes, and all possible
orientation(aligned/antialigned)-combinations of the binary components along
the $z-$ axis. We then perform the IMRCT with \qc templates (these being
routinely used by current implementations of the IMRCT), as well as eccentric
templates.

We find that, as expected, the IMRCT gets progressively biased with increasing
injected eccentricity while recovering with \qc templates. At $68\%$
confidence, we find that the IMRCT starts to get violated for eccentricities --
as defined by the waveform model used -- between $\eEOB \sim 0.05 - 0.09$ at a
reference frequency $\fref=20$ Hz, for our chosen CBC systems. Converting this
to a standardized definition of eccentricity \cite{Shaikh:2023ypz}, these
values become $\egw \sim 0.04 - 0.07$ at an orbital averaged frequency
$\avgfref=25$ Hz. Indeed, as the eccentricity is increased, the violations
occur even at $90\%$ confidence.

On the other hand, when recovering with eccentric templates, we find that, the
IMRCT is {\it always} satisfied, regardless of the spin magnitude, orientation,
or eccentricity value, at $68\%$ confidence. Our work therefore demonstrates
the need for accurate eccentric waveform models for IMRCT and possibly other
tests of GR. Moreover, our work complements and expands on previous work
studying the effect of neglect of eccentricity on the IMRCT. These either use
approximate GW PE likelihoods \cite{Bhat:2022amc}, or only study the potential
bias in the IMRCT caused by recovering a few eccentric numerical relativity
waveforms with phenomenological waveforms \cite{Narayan:2023vhm}.

Two important limitations of our work must be pointed out. The first is that
the injections are performed in zero-noise rather than real-noise. This could
potentially introduce additional fluctuations in the IMRCT as a consequence of
a particular noise realisation. The second is that the recovery of the IMRCT
has been tested using identical injection and recovery waveform models. It
would be of interest to consider eccentric numerical-relativity injections
recovered with other eccentric waveform models that are nominally used for GW
PE. Such a study might potentially demonstrate non-trivial biases in the IMRCT
even when eccentric waveforms are used, and if not, would ascertain the
feasibility of using current eccentric waveform models for tests of GR. We plan
to address the said limitations in future work.

\begin{acknowledgments}
We thank the anonymous referees for their useful comments and suggestions.
We thank Pankaj Saini for a careful reading of the draft. We also thank Prayush
Kumar, Nathan K. Johnson-McDaniel, Aditya Vijaykumar, and members of the
ICTS-TIFR Astrophysical Relativity Group, for useful discussions.
M.A.S.’s research was supported by the Department of Atomic Energy, Government
of India under project no. RTI4001 and the National Research Foundation of
Korea under grant No. NRF-2021R1A2C2012473. S.A.B.~acknowledges support from
the Department of Science and Technology and the Science and Engineering
Research Board (SERB) of India via Swarnajayanti Fellowship Grant
No.~DST/SJF/PSA-01/2017-18 and support from Infosys foundation. The authors
acknowledge the public code at
\url{https://github.com/osheamonn/eobresum_bilby/}, a modified version of which
was used as the source model for injection study with \bilby.  The authors are
grateful for computational resources provided by the LIGO Laboratory and
supported by National Science Foundation Grants PHY-0757058 and PHY-0823459.
The authors also acknowledge the use of the Alice cluster at the International
Centre for Theoretical Sciences.
\end{acknowledgments}

\bibliography{References}

\begin{thebibliography}{85}%
\makeatletter
\providecommand \@ifxundefined [1]{%
 \@ifx{#1\undefined}
}%
\providecommand \@ifnum [1]{%
 \ifnum #1\expandafter \@firstoftwo
 \else \expandafter \@secondoftwo
 \fi
}%
\providecommand \@ifx [1]{%
 \ifx #1\expandafter \@firstoftwo
 \else \expandafter \@secondoftwo
 \fi
}%
\providecommand \natexlab [1]{#1}%
\providecommand \enquote  [1]{``#1''}%
\providecommand \bibnamefont  [1]{#1}%
\providecommand \bibfnamefont [1]{#1}%
\providecommand \citenamefont [1]{#1}%
\providecommand \href@noop [0]{\@secondoftwo}%
\providecommand \href [0]{\begingroup \@sanitize@url \@href}%
\providecommand \@href[1]{\@@startlink{#1}\@@href}%
\providecommand \@@href[1]{\endgroup#1\@@endlink}%
\providecommand \@sanitize@url [0]{\catcode `\\12\catcode `\$12\catcode
  `\&12\catcode `\#12\catcode `\^12\catcode `\_12\catcode `\%12\relax}%
\providecommand \@@startlink[1]{}%
\providecommand \@@endlink[0]{}%
\providecommand \url  [0]{\begingroup\@sanitize@url \@url }%
\providecommand \@url [1]{\endgroup\@href {#1}{\urlprefix }}%
\providecommand \urlprefix  [0]{URL }%
\providecommand \Eprint [0]{\href }%
\providecommand \doibase [0]{http://dx.doi.org/}%
\providecommand \selectlanguage [0]{\@gobble}%
\providecommand \bibinfo  [0]{\@secondoftwo}%
\providecommand \bibfield  [0]{\@secondoftwo}%
\providecommand \translation [1]{[#1]}%
\providecommand \BibitemOpen [0]{}%
\providecommand \bibitemStop [0]{}%
\providecommand \bibitemNoStop [0]{.\EOS\space}%
\providecommand \EOS [0]{\spacefactor3000\relax}%
\providecommand \BibitemShut  [1]{\csname bibitem#1\endcsname}%
\let\auto@bib@innerbib\@empty
\bibitem [{\citenamefont {Aasi}\ \emph {et~al.}(2015)\citenamefont {Aasi} \emph
  {et~al.}}]{TheLIGOScientific:2014jea}%
  \BibitemOpen
  \bibfield  {author} {\bibinfo {author} {\bibfnamefont {J.}~\bibnamefont
  {Aasi}} \emph {et~al.} (\bibinfo {collaboration} {LIGO Scientific}),\
  }\bibfield  {title} {\enquote {\bibinfo {title} {{Advanced LIGO}},}\ }\href
  {\doibase 10.1088/0264-9381/32/7/074001} {\bibfield  {journal} {\bibinfo
  {journal} {Class. Quant. Grav.}\ }\textbf {\bibinfo {volume} {32}},\ \bibinfo
  {pages} {074001} (\bibinfo {year} {2015})},\ \Eprint
  {http://arxiv.org/abs/1411.4547} {arXiv:1411.4547 [gr-qc]} \BibitemShut
  {NoStop}%
\bibitem [{\citenamefont {Acernese}\ \emph {et~al.}(2015)\citenamefont
  {Acernese} \emph {et~al.}}]{TheVirgo:2014hva}%
  \BibitemOpen
  \bibfield  {author} {\bibinfo {author} {\bibfnamefont {F.}~\bibnamefont
  {Acernese}} \emph {et~al.} (\bibinfo {collaboration} {Virgo}),\ }\bibfield
  {title} {\enquote {\bibinfo {title} {{Advanced Virgo: a second-generation
  interferometric gravitational wave detector}},}\ }\href {\doibase
  10.1088/0264-9381/32/2/024001} {\bibfield  {journal} {\bibinfo  {journal}
  {Class. Quant. Grav.}\ }\textbf {\bibinfo {volume} {32}},\ \bibinfo {pages}
  {024001} (\bibinfo {year} {2015})},\ \Eprint {http://arxiv.org/abs/1408.3978}
  {arXiv:1408.3978 [gr-qc]} \BibitemShut {NoStop}%
\bibitem [{\citenamefont {Abbott}\ \emph {et~al.}(2023)\citenamefont {Abbott}
  \emph {et~al.}}]{KAGRA:2021vkt}%
  \BibitemOpen
  \bibfield  {author} {\bibinfo {author} {\bibfnamefont {R.}~\bibnamefont
  {Abbott}} \emph {et~al.} (\bibinfo {collaboration} {KAGRA, VIRGO, LIGO
  Scientific}),\ }\bibfield  {title} {\enquote {\bibinfo {title} {{GWTC-3:
  Compact Binary Coalescences Observed by LIGO and Virgo during the Second Part
  of the Third Observing Run}},}\ }\href {\doibase 10.1103/PhysRevX.13.041039}
  {\bibfield  {journal} {\bibinfo  {journal} {Phys. Rev. X}\ }\textbf {\bibinfo
  {volume} {13}},\ \bibinfo {pages} {041039} (\bibinfo {year} {2023})},\
  \Eprint {http://arxiv.org/abs/2111.03606} {arXiv:2111.03606 [gr-qc]}
  \BibitemShut {NoStop}%
\bibitem [{\citenamefont {Abbott}\ \emph {et~al.}(2017)\citenamefont {Abbott}
  \emph {et~al.}}]{TheLIGOScientific:2017qsa}%
  \BibitemOpen
  \bibfield  {author} {\bibinfo {author} {\bibfnamefont {Benjamin~P.}\
  \bibnamefont {Abbott}} \emph {et~al.} (\bibinfo {collaboration} {LIGO
  Scientific, Virgo}),\ }\bibfield  {title} {\enquote {\bibinfo {title}
  {{GW170817: Observation of Gravitational Waves from a Binary Neutron Star
  Inspiral}},}\ }\href {\doibase 10.1103/PhysRevLett.119.161101} {\bibfield
  {journal} {\bibinfo  {journal} {Phys. Rev. Lett.}\ }\textbf {\bibinfo
  {volume} {119}},\ \bibinfo {pages} {161101} (\bibinfo {year} {2017})},\
  \Eprint {http://arxiv.org/abs/1710.05832} {arXiv:1710.05832 [gr-qc]}
  \BibitemShut {NoStop}%
\bibitem [{\citenamefont {Abbott}\ \emph
  {et~al.}(2020{\natexlab{a}})\citenamefont {Abbott} \emph
  {et~al.}}]{Abbott:2020uma}%
  \BibitemOpen
  \bibfield  {author} {\bibinfo {author} {\bibfnamefont {B.P.}\ \bibnamefont
  {Abbott}} \emph {et~al.} (\bibinfo {collaboration} {LIGO Scientific,
  Virgo}),\ }\bibfield  {title} {\enquote {\bibinfo {title} {{GW190425:
  Observation of a Compact Binary Coalescence with Total Mass $\sim 3.4
  M_{\odot}$}},}\ }\href {\doibase 10.3847/2041-8213/ab75f5} {\bibfield
  {journal} {\bibinfo  {journal} {Astrophys. J. Lett.}\ }\textbf {\bibinfo
  {volume} {892}},\ \bibinfo {pages} {L3} (\bibinfo {year}
  {2020}{\natexlab{a}})},\ \Eprint {http://arxiv.org/abs/2001.01761}
  {arXiv:2001.01761 [astro-ph.HE]} \BibitemShut {NoStop}%
\bibitem [{\citenamefont {Abbott}\ \emph
  {et~al.}(2021{\natexlab{a}})\citenamefont {Abbott} \emph
  {et~al.}}]{LIGOScientific:2021qlt}%
  \BibitemOpen
  \bibfield  {author} {\bibinfo {author} {\bibfnamefont {R.}~\bibnamefont
  {Abbott}} \emph {et~al.} (\bibinfo {collaboration} {LIGO Scientific, KAGRA,
  VIRGO}),\ }\bibfield  {title} {\enquote {\bibinfo {title} {{Observation of
  Gravitational Waves from Two Neutron Star\textendash{}Black Hole
  Coalescences}},}\ }\href {\doibase 10.3847/2041-8213/ac082e} {\bibfield
  {journal} {\bibinfo  {journal} {Astrophys. J. Lett.}\ }\textbf {\bibinfo
  {volume} {915}},\ \bibinfo {pages} {L5} (\bibinfo {year}
  {2021}{\natexlab{a}})},\ \Eprint {http://arxiv.org/abs/2106.15163}
  {arXiv:2106.15163 [astro-ph.HE]} \BibitemShut {NoStop}%
\bibitem [{\citenamefont {Sathyaprakash}\ and\ \citenamefont
  {Schutz}(2009)}]{Sathyaprakash:2009xs}%
  \BibitemOpen
  \bibfield  {author} {\bibinfo {author} {\bibfnamefont {B.~S.}\ \bibnamefont
  {Sathyaprakash}}\ and\ \bibinfo {author} {\bibfnamefont {B.~F.}\ \bibnamefont
  {Schutz}},\ }\bibfield  {title} {\enquote {\bibinfo {title} {{Physics,
  Astrophysics and Cosmology with Gravitational Waves}},}\ }\href {\doibase
  10.12942/lrr-2009-2} {\bibfield  {journal} {\bibinfo  {journal} {Living Rev.
  Rel.}\ }\textbf {\bibinfo {volume} {12}},\ \bibinfo {pages} {2} (\bibinfo
  {year} {2009})},\ \Eprint {http://arxiv.org/abs/0903.0338} {arXiv:0903.0338
  [gr-qc]} \BibitemShut {NoStop}%
\bibitem [{\citenamefont {Akiyama}\ \emph {et~al.}(2019)\citenamefont {Akiyama}
  \emph {et~al.}}]{EventHorizonTelescope:2019dse}%
  \BibitemOpen
  \bibfield  {author} {\bibinfo {author} {\bibfnamefont {Kazunori}\
  \bibnamefont {Akiyama}} \emph {et~al.} (\bibinfo {collaboration} {Event
  Horizon Telescope}),\ }\bibfield  {title} {\enquote {\bibinfo {title} {{First
  M87 Event Horizon Telescope Results. I. The Shadow of the Supermassive Black
  Hole}},}\ }\href {\doibase 10.3847/2041-8213/ab0ec7} {\bibfield  {journal}
  {\bibinfo  {journal} {Astrophys. J. Lett.}\ }\textbf {\bibinfo {volume}
  {875}},\ \bibinfo {pages} {L1} (\bibinfo {year} {2019})},\ \Eprint
  {http://arxiv.org/abs/1906.11238} {arXiv:1906.11238 [astro-ph.GA]}
  \BibitemShut {NoStop}%
\bibitem [{\citenamefont {Psaltis}(2019)}]{Psaltis:2018xkc}%
  \BibitemOpen
  \bibfield  {author} {\bibinfo {author} {\bibfnamefont {Dimitrios}\
  \bibnamefont {Psaltis}},\ }\bibfield  {title} {\enquote {\bibinfo {title}
  {{Testing General Relativity with the Event Horizon Telescope}},}\ }\href
  {\doibase 10.1007/s10714-019-2611-5} {\bibfield  {journal} {\bibinfo
  {journal} {Gen. Rel. Grav.}\ }\textbf {\bibinfo {volume} {51}},\ \bibinfo
  {pages} {137} (\bibinfo {year} {2019})},\ \Eprint
  {http://arxiv.org/abs/1806.09740} {arXiv:1806.09740 [astro-ph.HE]}
  \BibitemShut {NoStop}%
\bibitem [{\citenamefont {Akutsu}\ \emph {et~al.}(2021)\citenamefont {Akutsu}
  \emph {et~al.}}]{KAGRA:2020tym}%
  \BibitemOpen
  \bibfield  {author} {\bibinfo {author} {\bibfnamefont {T.}~\bibnamefont
  {Akutsu}} \emph {et~al.} (\bibinfo {collaboration} {KAGRA}),\ }\bibfield
  {title} {\enquote {\bibinfo {title} {{Overview of KAGRA: Detector design and
  construction history}},}\ }\href {\doibase 10.1093/ptep/ptaa125} {\bibfield
  {journal} {\bibinfo  {journal} {PTEP}\ }\textbf {\bibinfo {volume} {2021}},\
  \bibinfo {pages} {05A101} (\bibinfo {year} {2021})},\ \Eprint
  {http://arxiv.org/abs/2005.05574} {arXiv:2005.05574 [physics.ins-det]}
  \BibitemShut {NoStop}%
\bibitem [{\citenamefont {Abbott}\ \emph
  {et~al.}(2021{\natexlab{b}})\citenamefont {Abbott} \emph
  {et~al.}}]{LIGOScientific:2021sio}%
  \BibitemOpen
  \bibfield  {author} {\bibinfo {author} {\bibfnamefont {R.}~\bibnamefont
  {Abbott}} \emph {et~al.} (\bibinfo {collaboration} {LIGO Scientific, VIRGO,
  KAGRA}),\ }\bibfield  {title} {\enquote {\bibinfo {title} {{Tests of General
  Relativity with GWTC-3}},}\ }\href@noop {} {\  (\bibinfo {year}
  {2021}{\natexlab{b}})},\ \Eprint {http://arxiv.org/abs/2112.06861}
  {arXiv:2112.06861 [gr-qc]} \BibitemShut {NoStop}%
\bibitem [{\citenamefont {Abbott}\ \emph {et~al.}(2016)\citenamefont {Abbott}
  \emph {et~al.}}]{LIGOScientific:2016lio}%
  \BibitemOpen
  \bibfield  {author} {\bibinfo {author} {\bibfnamefont {B.~P.}\ \bibnamefont
  {Abbott}} \emph {et~al.} (\bibinfo {collaboration} {LIGO Scientific,
  Virgo}),\ }\bibfield  {title} {\enquote {\bibinfo {title} {{Tests of general
  relativity with GW150914}},}\ }\href {\doibase
  10.1103/PhysRevLett.116.221101} {\bibfield  {journal} {\bibinfo  {journal}
  {Phys. Rev. Lett.}\ }\textbf {\bibinfo {volume} {116}},\ \bibinfo {pages}
  {221101} (\bibinfo {year} {2016})},\ \bibinfo {note} {[Erratum:
  Phys.Rev.Lett. 121, 129902 (2018)]},\ \Eprint
  {http://arxiv.org/abs/1602.03841} {arXiv:1602.03841 [gr-qc]} \BibitemShut
  {NoStop}%
\bibitem [{\citenamefont {Abbott}\ \emph
  {et~al.}(2019{\natexlab{a}})\citenamefont {Abbott} \emph
  {et~al.}}]{LIGOScientific:2019fpa}%
  \BibitemOpen
  \bibfield  {author} {\bibinfo {author} {\bibfnamefont {B.~P.}\ \bibnamefont
  {Abbott}} \emph {et~al.} (\bibinfo {collaboration} {LIGO Scientific,
  Virgo}),\ }\bibfield  {title} {\enquote {\bibinfo {title} {{Tests of General
  Relativity with the Binary Black Hole Signals from the LIGO-Virgo Catalog
  GWTC-1}},}\ }\href {\doibase 10.1103/PhysRevD.100.104036} {\bibfield
  {journal} {\bibinfo  {journal} {Phys. Rev. D}\ }\textbf {\bibinfo {volume}
  {100}},\ \bibinfo {pages} {104036} (\bibinfo {year} {2019}{\natexlab{a}})},\
  \Eprint {http://arxiv.org/abs/1903.04467} {arXiv:1903.04467 [gr-qc]}
  \BibitemShut {NoStop}%
\bibitem [{\citenamefont {Abbott}\ \emph
  {et~al.}(2020{\natexlab{b}})\citenamefont {Abbott} \emph
  {et~al.}}]{LIGOScientific:2020ufj}%
  \BibitemOpen
  \bibfield  {author} {\bibinfo {author} {\bibfnamefont {R.}~\bibnamefont
  {Abbott}} \emph {et~al.} (\bibinfo {collaboration} {LIGO Scientific,
  Virgo}),\ }\bibfield  {title} {\enquote {\bibinfo {title} {{Properties and
  Astrophysical Implications of the 150 M$_\odot$ Binary Black Hole Merger
  GW190521}},}\ }\href {\doibase 10.3847/2041-8213/aba493} {\bibfield
  {journal} {\bibinfo  {journal} {Astrophys. J. Lett.}\ }\textbf {\bibinfo
  {volume} {900}},\ \bibinfo {pages} {L13} (\bibinfo {year}
  {2020}{\natexlab{b}})},\ \Eprint {http://arxiv.org/abs/2009.01190}
  {arXiv:2009.01190 [astro-ph.HE]} \BibitemShut {NoStop}%
\bibitem [{\citenamefont {Blanchet}\ and\ \citenamefont
  {Sathyaprakash}(1995)}]{Blanchet:1994ez}%
  \BibitemOpen
  \bibfield  {author} {\bibinfo {author} {\bibfnamefont {Luc}\ \bibnamefont
  {Blanchet}}\ and\ \bibinfo {author} {\bibfnamefont {B.~S.}\ \bibnamefont
  {Sathyaprakash}},\ }\bibfield  {title} {\enquote {\bibinfo {title}
  {{Detecting the tail effect in gravitational wave experiments}},}\ }\href
  {\doibase 10.1103/PhysRevLett.74.1067} {\bibfield  {journal} {\bibinfo
  {journal} {Phys. Rev. Lett.}\ }\textbf {\bibinfo {volume} {74}},\ \bibinfo
  {pages} {1067--1070} (\bibinfo {year} {1995})}\BibitemShut {NoStop}%
\bibitem [{\citenamefont {Blanchet}\ and\ \citenamefont
  {Sathyaprakash}(1994)}]{blanchet1994signal}%
  \BibitemOpen
  \bibfield  {author} {\bibinfo {author} {\bibfnamefont {Luc}\ \bibnamefont
  {Blanchet}}\ and\ \bibinfo {author} {\bibfnamefont {Bangalore~Suryanarayana}\
  \bibnamefont {Sathyaprakash}},\ }\bibfield  {title} {\enquote {\bibinfo
  {title} {Signal analysis of gravitational wave tails},}\ }\href@noop {}
  {\bibfield  {journal} {\bibinfo  {journal} {Classical and Quantum Gravity}\
  }\textbf {\bibinfo {volume} {11}},\ \bibinfo {pages} {2807} (\bibinfo {year}
  {1994})}\BibitemShut {NoStop}%
\bibitem [{\citenamefont {Arun}\ \emph
  {et~al.}(2006{\natexlab{a}})\citenamefont {Arun}, \citenamefont {Iyer},
  \citenamefont {Qusailah},\ and\ \citenamefont {Sathyaprakash}}]{Arun:2006hn}%
  \BibitemOpen
  \bibfield  {author} {\bibinfo {author} {\bibfnamefont {K.~G.}\ \bibnamefont
  {Arun}}, \bibinfo {author} {\bibfnamefont {Bala~R.}\ \bibnamefont {Iyer}},
  \bibinfo {author} {\bibfnamefont {M.~S.~S.}\ \bibnamefont {Qusailah}}, \ and\
  \bibinfo {author} {\bibfnamefont {B.~S.}\ \bibnamefont {Sathyaprakash}},\
  }\bibfield  {title} {\enquote {\bibinfo {title} {{Probing the non-linear
  structure of general relativity with black hole binaries}},}\ }\href
  {\doibase 10.1103/PhysRevD.74.024006} {\bibfield  {journal} {\bibinfo
  {journal} {Phys. Rev. D}\ }\textbf {\bibinfo {volume} {74}},\ \bibinfo
  {pages} {024006} (\bibinfo {year} {2006}{\natexlab{a}})},\ \Eprint
  {http://arxiv.org/abs/gr-qc/0604067} {arXiv:gr-qc/0604067} \BibitemShut
  {NoStop}%
\bibitem [{\citenamefont {Arun}\ \emph
  {et~al.}(2006{\natexlab{b}})\citenamefont {Arun}, \citenamefont {Iyer},
  \citenamefont {Qusailah},\ and\ \citenamefont {Sathyaprakash}}]{Arun:2006yw}%
  \BibitemOpen
  \bibfield  {author} {\bibinfo {author} {\bibfnamefont {K.~G.}\ \bibnamefont
  {Arun}}, \bibinfo {author} {\bibfnamefont {Bala~R.}\ \bibnamefont {Iyer}},
  \bibinfo {author} {\bibfnamefont {M.~S.~S.}\ \bibnamefont {Qusailah}}, \ and\
  \bibinfo {author} {\bibfnamefont {B.~S.}\ \bibnamefont {Sathyaprakash}},\
  }\bibfield  {title} {\enquote {\bibinfo {title} {{Testing post-Newtonian
  theory with gravitational wave observations}},}\ }\href {\doibase
  10.1088/0264-9381/23/9/L01} {\bibfield  {journal} {\bibinfo  {journal}
  {Class. Quant. Grav.}\ }\textbf {\bibinfo {volume} {23}},\ \bibinfo {pages}
  {L37--L43} (\bibinfo {year} {2006}{\natexlab{b}})},\ \Eprint
  {http://arxiv.org/abs/gr-qc/0604018} {arXiv:gr-qc/0604018} \BibitemShut
  {NoStop}%
\bibitem [{\citenamefont {Yunes}\ and\ \citenamefont
  {Pretorius}(2009)}]{Yunes:2009ke}%
  \BibitemOpen
  \bibfield  {author} {\bibinfo {author} {\bibfnamefont {Nicolas}\ \bibnamefont
  {Yunes}}\ and\ \bibinfo {author} {\bibfnamefont {Frans}\ \bibnamefont
  {Pretorius}},\ }\bibfield  {title} {\enquote {\bibinfo {title} {{Fundamental
  Theoretical Bias in Gravitational Wave Astrophysics and the Parameterized
  Post-Einsteinian Framework}},}\ }\href {\doibase 10.1103/PhysRevD.80.122003}
  {\bibfield  {journal} {\bibinfo  {journal} {Phys. Rev. D}\ }\textbf {\bibinfo
  {volume} {80}},\ \bibinfo {pages} {122003} (\bibinfo {year} {2009})},\
  \Eprint {http://arxiv.org/abs/0909.3328} {arXiv:0909.3328 [gr-qc]}
  \BibitemShut {NoStop}%
\bibitem [{\citenamefont {Mishra}\ \emph {et~al.}(2010)\citenamefont {Mishra},
  \citenamefont {Arun}, \citenamefont {Iyer},\ and\ \citenamefont
  {Sathyaprakash}}]{Mishra:2010tp}%
  \BibitemOpen
  \bibfield  {author} {\bibinfo {author} {\bibfnamefont {Chandra~Kant}\
  \bibnamefont {Mishra}}, \bibinfo {author} {\bibfnamefont {K.~G.}\
  \bibnamefont {Arun}}, \bibinfo {author} {\bibfnamefont {Bala~R.}\
  \bibnamefont {Iyer}}, \ and\ \bibinfo {author} {\bibfnamefont {B.~S.}\
  \bibnamefont {Sathyaprakash}},\ }\bibfield  {title} {\enquote {\bibinfo
  {title} {{Parametrized tests of post-Newtonian theory using Advanced LIGO and
  Einstein Telescope}},}\ }\href {\doibase 10.1103/PhysRevD.82.064010}
  {\bibfield  {journal} {\bibinfo  {journal} {Phys. Rev. D}\ }\textbf {\bibinfo
  {volume} {82}},\ \bibinfo {pages} {064010} (\bibinfo {year} {2010})},\
  \Eprint {http://arxiv.org/abs/1005.0304} {arXiv:1005.0304 [gr-qc]}
  \BibitemShut {NoStop}%
\bibitem [{\citenamefont {Li}\ \emph {et~al.}(2012{\natexlab{a}})\citenamefont
  {Li}, \citenamefont {Del~Pozzo}, \citenamefont {Vitale}, \citenamefont {Van
  Den~Broeck}, \citenamefont {Agathos}, \citenamefont {Veitch}, \citenamefont
  {Grover}, \citenamefont {Sidery}, \citenamefont {Sturani},\ and\
  \citenamefont {Vecchio}}]{Li:2011cg}%
  \BibitemOpen
  \bibfield  {author} {\bibinfo {author} {\bibfnamefont {T.~G.~F.}\
  \bibnamefont {Li}}, \bibinfo {author} {\bibfnamefont {W.}~\bibnamefont
  {Del~Pozzo}}, \bibinfo {author} {\bibfnamefont {S.}~\bibnamefont {Vitale}},
  \bibinfo {author} {\bibfnamefont {C.}~\bibnamefont {Van Den~Broeck}},
  \bibinfo {author} {\bibfnamefont {M.}~\bibnamefont {Agathos}}, \bibinfo
  {author} {\bibfnamefont {J.}~\bibnamefont {Veitch}}, \bibinfo {author}
  {\bibfnamefont {K.}~\bibnamefont {Grover}}, \bibinfo {author} {\bibfnamefont
  {T.}~\bibnamefont {Sidery}}, \bibinfo {author} {\bibfnamefont
  {R.}~\bibnamefont {Sturani}}, \ and\ \bibinfo {author} {\bibfnamefont
  {A.}~\bibnamefont {Vecchio}},\ }\bibfield  {title} {\enquote {\bibinfo
  {title} {{Towards a generic test of the strong field dynamics of general
  relativity using compact binary coalescence}},}\ }\href {\doibase
  10.1103/PhysRevD.85.082003} {\bibfield  {journal} {\bibinfo  {journal} {Phys.
  Rev. D}\ }\textbf {\bibinfo {volume} {85}},\ \bibinfo {pages} {082003}
  (\bibinfo {year} {2012}{\natexlab{a}})},\ \Eprint
  {http://arxiv.org/abs/1110.0530} {arXiv:1110.0530 [gr-qc]} \BibitemShut
  {NoStop}%
\bibitem [{\citenamefont {Li}\ \emph {et~al.}(2012{\natexlab{b}})\citenamefont
  {Li}, \citenamefont {Del~Pozzo}, \citenamefont {Vitale}, \citenamefont {Van
  Den~Broeck}, \citenamefont {Agathos}, \citenamefont {Veitch}, \citenamefont
  {Grover}, \citenamefont {Sidery}, \citenamefont {Sturani},\ and\
  \citenamefont {Vecchio}}]{Li:2011vx}%
  \BibitemOpen
  \bibfield  {author} {\bibinfo {author} {\bibfnamefont {T.~G.~F.}\
  \bibnamefont {Li}}, \bibinfo {author} {\bibfnamefont {W.}~\bibnamefont
  {Del~Pozzo}}, \bibinfo {author} {\bibfnamefont {S.}~\bibnamefont {Vitale}},
  \bibinfo {author} {\bibfnamefont {C.}~\bibnamefont {Van Den~Broeck}},
  \bibinfo {author} {\bibfnamefont {M.}~\bibnamefont {Agathos}}, \bibinfo
  {author} {\bibfnamefont {J.}~\bibnamefont {Veitch}}, \bibinfo {author}
  {\bibfnamefont {K.}~\bibnamefont {Grover}}, \bibinfo {author} {\bibfnamefont
  {T.}~\bibnamefont {Sidery}}, \bibinfo {author} {\bibfnamefont
  {R.}~\bibnamefont {Sturani}}, \ and\ \bibinfo {author} {\bibfnamefont
  {A.}~\bibnamefont {Vecchio}},\ }\bibfield  {title} {\enquote {\bibinfo
  {title} {{Towards a generic test of the strong field dynamics of general
  relativity using compact binary coalescence: Further investigations}},}\
  }\href {\doibase 10.1088/1742-6596/363/1/012028} {\bibfield  {journal}
  {\bibinfo  {journal} {J. Phys. Conf. Ser.}\ }\textbf {\bibinfo {volume}
  {363}},\ \bibinfo {pages} {012028} (\bibinfo {year} {2012}{\natexlab{b}})},\
  \Eprint {http://arxiv.org/abs/1111.5274} {arXiv:1111.5274 [gr-qc]}
  \BibitemShut {NoStop}%
\bibitem [{\citenamefont {Ghosh}\ \emph {et~al.}(2016)\citenamefont {Ghosh}
  \emph {et~al.}}]{Ghosh:2016qgn}%
  \BibitemOpen
  \bibfield  {author} {\bibinfo {author} {\bibfnamefont {Abhirup}\ \bibnamefont
  {Ghosh}} \emph {et~al.},\ }\bibfield  {title} {\enquote {\bibinfo {title}
  {{Testing general relativity using golden black-hole binaries}},}\ }\href
  {\doibase 10.1103/PhysRevD.94.021101} {\bibfield  {journal} {\bibinfo
  {journal} {Phys. Rev. D}\ }\textbf {\bibinfo {volume} {94}},\ \bibinfo
  {pages} {021101} (\bibinfo {year} {2016})},\ \Eprint
  {http://arxiv.org/abs/1602.02453} {arXiv:1602.02453 [gr-qc]} \BibitemShut
  {NoStop}%
\bibitem [{\citenamefont {Ghosh}\ \emph {et~al.}(2018)\citenamefont {Ghosh},
  \citenamefont {Johnson-Mcdaniel}, \citenamefont {Ghosh}, \citenamefont
  {Mishra}, \citenamefont {Ajith}, \citenamefont {Del~Pozzo}, \citenamefont
  {Berry}, \citenamefont {Nielsen},\ and\ \citenamefont
  {London}}]{Ghosh:2017gfp}%
  \BibitemOpen
  \bibfield  {author} {\bibinfo {author} {\bibfnamefont {Abhirup}\ \bibnamefont
  {Ghosh}}, \bibinfo {author} {\bibfnamefont {Nathan~K.}\ \bibnamefont
  {Johnson-Mcdaniel}}, \bibinfo {author} {\bibfnamefont {Archisman}\
  \bibnamefont {Ghosh}}, \bibinfo {author} {\bibfnamefont {Chandra~Kant}\
  \bibnamefont {Mishra}}, \bibinfo {author} {\bibfnamefont {Parameswaran}\
  \bibnamefont {Ajith}}, \bibinfo {author} {\bibfnamefont {Walter}\
  \bibnamefont {Del~Pozzo}}, \bibinfo {author} {\bibfnamefont {Christopher
  P.~L.}\ \bibnamefont {Berry}}, \bibinfo {author} {\bibfnamefont {Alex~B.}\
  \bibnamefont {Nielsen}}, \ and\ \bibinfo {author} {\bibfnamefont {Lionel}\
  \bibnamefont {London}},\ }\bibfield  {title} {\enquote {\bibinfo {title}
  {{Testing general relativity using gravitational wave signals from the
  inspiral, merger and ringdown of binary black holes}},}\ }\href {\doibase
  10.1088/1361-6382/aa972e} {\bibfield  {journal} {\bibinfo  {journal} {Class.
  Quant. Grav.}\ }\textbf {\bibinfo {volume} {35}},\ \bibinfo {pages} {014002}
  (\bibinfo {year} {2018})},\ \Eprint {http://arxiv.org/abs/1704.06784}
  {arXiv:1704.06784 [gr-qc]} \BibitemShut {NoStop}%
\bibitem [{\citenamefont {Abbott}\ \emph
  {et~al.}(2019{\natexlab{b}})\citenamefont {Abbott} \emph
  {et~al.}}]{LIGOScientific:2018dkp}%
  \BibitemOpen
  \bibfield  {author} {\bibinfo {author} {\bibfnamefont {B.~P.}\ \bibnamefont
  {Abbott}} \emph {et~al.} (\bibinfo {collaboration} {LIGO Scientific,
  Virgo}),\ }\bibfield  {title} {\enquote {\bibinfo {title} {{Tests of General
  Relativity with GW170817}},}\ }\href {\doibase
  10.1103/PhysRevLett.123.011102} {\bibfield  {journal} {\bibinfo  {journal}
  {Phys. Rev. Lett.}\ }\textbf {\bibinfo {volume} {123}},\ \bibinfo {pages}
  {011102} (\bibinfo {year} {2019}{\natexlab{b}})},\ \Eprint
  {http://arxiv.org/abs/1811.00364} {arXiv:1811.00364 [gr-qc]} \BibitemShut
  {NoStop}%
\bibitem [{\citenamefont {Will}(1998)}]{Will:1997bb}%
  \BibitemOpen
  \bibfield  {author} {\bibinfo {author} {\bibfnamefont {Clifford~M.}\
  \bibnamefont {Will}},\ }\bibfield  {title} {\enquote {\bibinfo {title}
  {{Bounding the mass of the graviton using gravitational wave observations of
  inspiralling compact binaries}},}\ }\href {\doibase 10.1103/PhysRevD.57.2061}
  {\bibfield  {journal} {\bibinfo  {journal} {Phys. Rev. D}\ }\textbf {\bibinfo
  {volume} {57}},\ \bibinfo {pages} {2061--2068} (\bibinfo {year} {1998})},\
  \Eprint {http://arxiv.org/abs/gr-qc/9709011} {arXiv:gr-qc/9709011}
  \BibitemShut {NoStop}%
\bibitem [{\citenamefont {Peters}\ and\ \citenamefont
  {Mathews}(1963)}]{Peters:1963ux}%
  \BibitemOpen
  \bibfield  {author} {\bibinfo {author} {\bibfnamefont {P.~C.}\ \bibnamefont
  {Peters}}\ and\ \bibinfo {author} {\bibfnamefont {J.}~\bibnamefont
  {Mathews}},\ }\bibfield  {title} {\enquote {\bibinfo {title} {{Gravitational
  radiation from point masses in a Keplerian orbit}},}\ }\href {\doibase
  10.1103/PhysRev.131.435} {\bibfield  {journal} {\bibinfo  {journal} {Phys.
  Rev.}\ }\textbf {\bibinfo {volume} {131}},\ \bibinfo {pages} {435--439}
  (\bibinfo {year} {1963})}\BibitemShut {NoStop}%
\bibitem [{\citenamefont {Peters}(1964)}]{Peters:1964zz}%
  \BibitemOpen
  \bibfield  {author} {\bibinfo {author} {\bibfnamefont {P.~C.}\ \bibnamefont
  {Peters}},\ }\bibfield  {title} {\enquote {\bibinfo {title} {{Gravitational
  Radiation and the Motion of Two Point Masses}},}\ }\href {\doibase
  10.1103/PhysRev.136.B1224} {\bibfield  {journal} {\bibinfo  {journal} {Phys.
  Rev.}\ }\textbf {\bibinfo {volume} {136}},\ \bibinfo {pages} {B1224--B1232}
  (\bibinfo {year} {1964})}\BibitemShut {NoStop}%
\bibitem [{\citenamefont {Abac}\ \emph {et~al.}(2023)\citenamefont {Abac} \emph
  {et~al.}}]{LIGOScientific:2023lpe}%
  \BibitemOpen
  \bibfield  {author} {\bibinfo {author} {\bibfnamefont {A.~G.}\ \bibnamefont
  {Abac}} \emph {et~al.} (\bibinfo {collaboration} {LIGO Scientific, VIRGO,
  KAGRA}),\ }\bibfield  {title} {\enquote {\bibinfo {title} {{Search for
  Eccentric Black Hole Coalescences during the Third Observing Run of LIGO and
  Virgo}},}\ }\href@noop {} {\  (\bibinfo {year} {2023})},\ \Eprint
  {http://arxiv.org/abs/2308.03822} {arXiv:2308.03822 [astro-ph.HE]}
  \BibitemShut {NoStop}%
\bibitem [{\citenamefont {Abbott}\ \emph
  {et~al.}(2019{\natexlab{c}})\citenamefont {Abbott} \emph
  {et~al.}}]{LIGOScientific:2019dag}%
  \BibitemOpen
  \bibfield  {author} {\bibinfo {author} {\bibfnamefont {B.~P.}\ \bibnamefont
  {Abbott}} \emph {et~al.} (\bibinfo {collaboration} {LIGO Scientific,
  Virgo}),\ }\bibfield  {title} {\enquote {\bibinfo {title} {{Search for
  Eccentric Binary Black Hole Mergers with Advanced LIGO and Advanced Virgo
  during their First and Second Observing Runs}},}\ }\href {\doibase
  10.3847/1538-4357/ab3c2d} {\bibfield  {journal} {\bibinfo  {journal}
  {Astrophys. J.}\ }\textbf {\bibinfo {volume} {883}},\ \bibinfo {pages} {149}
  (\bibinfo {year} {2019}{\natexlab{c}})},\ \Eprint
  {http://arxiv.org/abs/1907.09384} {arXiv:1907.09384 [astro-ph.HE]}
  \BibitemShut {NoStop}%
\bibitem [{\citenamefont {Nitz}\ \emph {et~al.}(2019)\citenamefont {Nitz},
  \citenamefont {Lenon},\ and\ \citenamefont {Brown}}]{Nitz:2019spj}%
  \BibitemOpen
  \bibfield  {author} {\bibinfo {author} {\bibfnamefont {Alexander~H.}\
  \bibnamefont {Nitz}}, \bibinfo {author} {\bibfnamefont {Amber}\ \bibnamefont
  {Lenon}}, \ and\ \bibinfo {author} {\bibfnamefont {Duncan~A.}\ \bibnamefont
  {Brown}},\ }\bibfield  {title} {\enquote {\bibinfo {title} {{Search for
  Eccentric Binary Neutron Star Mergers in the first and second observing runs
  of Advanced LIGO}},}\ }\href {\doibase 10.3847/1538-4357/ab6611} {\bibfield
  {journal} {\bibinfo  {journal} {Astrophys. J.}\ }\textbf {\bibinfo {volume}
  {890}},\ \bibinfo {pages} {1} (\bibinfo {year} {2019})},\ \Eprint
  {http://arxiv.org/abs/1912.05464} {arXiv:1912.05464 [astro-ph.HE]}
  \BibitemShut {NoStop}%
\bibitem [{\citenamefont {{Mapelli}}(2020)}]{Mapelli:2021for}%
  \BibitemOpen
  \bibfield  {author} {\bibinfo {author} {\bibfnamefont {Michela}\ \bibnamefont
  {{Mapelli}}},\ }\bibfield  {title} {\enquote {\bibinfo {title} {{Binary black
  hole mergers: formation and populations}},}\ }\href {\doibase
  10.3389/fspas.2020.00038} {\bibfield  {journal} {\bibinfo  {journal}
  {Frontiers in Astronomy and Space Sciences}\ }\textbf {\bibinfo {volume}
  {7}},\ \bibinfo {eid} {38} (\bibinfo {year} {2020})},\ \Eprint
  {http://arxiv.org/abs/2105.12455} {arXiv:2105.12455 [astro-ph.HE]}
  \BibitemShut {NoStop}%
\bibitem [{\citenamefont {Samsing}(2018)}]{Samsing:2017xmd}%
  \BibitemOpen
  \bibfield  {author} {\bibinfo {author} {\bibfnamefont {Johan}\ \bibnamefont
  {Samsing}},\ }\bibfield  {title} {\enquote {\bibinfo {title} {{Eccentric
  Black Hole Mergers Forming in Globular Clusters}},}\ }\href {\doibase
  10.1103/PhysRevD.97.103014} {\bibfield  {journal} {\bibinfo  {journal} {Phys.
  Rev. D}\ }\textbf {\bibinfo {volume} {97}},\ \bibinfo {pages} {103014}
  (\bibinfo {year} {2018})},\ \Eprint {http://arxiv.org/abs/1711.07452}
  {arXiv:1711.07452 [astro-ph.HE]} \BibitemShut {NoStop}%
\bibitem [{\citenamefont {Zevin}\ \emph {et~al.}(2019)\citenamefont {Zevin},
  \citenamefont {Samsing}, \citenamefont {Rodriguez}, \citenamefont {Haster},\
  and\ \citenamefont {Ramirez-Ruiz}}]{Zevin:2018kzq}%
  \BibitemOpen
  \bibfield  {author} {\bibinfo {author} {\bibfnamefont {Michael}\ \bibnamefont
  {Zevin}}, \bibinfo {author} {\bibfnamefont {Johan}\ \bibnamefont {Samsing}},
  \bibinfo {author} {\bibfnamefont {Carl}\ \bibnamefont {Rodriguez}}, \bibinfo
  {author} {\bibfnamefont {Carl-Johan}\ \bibnamefont {Haster}}, \ and\ \bibinfo
  {author} {\bibfnamefont {Enrico}\ \bibnamefont {Ramirez-Ruiz}},\ }\bibfield
  {title} {\enquote {\bibinfo {title} {{Eccentric Black Hole Mergers in Dense
  Star Clusters: The Role of Binary\textendash{}Binary Encounters}},}\ }\href
  {\doibase 10.3847/1538-4357/aaf6ec} {\bibfield  {journal} {\bibinfo
  {journal} {Astrophys. J.}\ }\textbf {\bibinfo {volume} {871}},\ \bibinfo
  {pages} {91} (\bibinfo {year} {2019})},\ \Eprint
  {http://arxiv.org/abs/1810.00901} {arXiv:1810.00901 [astro-ph.HE]}
  \BibitemShut {NoStop}%
\bibitem [{\citenamefont {Romero-Shaw}\ \emph {et~al.}(2022)\citenamefont
  {Romero-Shaw}, \citenamefont {Lasky},\ and\ \citenamefont
  {Thrane}}]{Romero-Shaw:2022xko}%
  \BibitemOpen
  \bibfield  {author} {\bibinfo {author} {\bibfnamefont {Isobel~M.}\
  \bibnamefont {Romero-Shaw}}, \bibinfo {author} {\bibfnamefont {Paul~D.}\
  \bibnamefont {Lasky}}, \ and\ \bibinfo {author} {\bibfnamefont {Eric}\
  \bibnamefont {Thrane}},\ }\bibfield  {title} {\enquote {\bibinfo {title}
  {{Four Eccentric Mergers Increase the Evidence that
  LIGO\textendash{}Virgo\textendash{}KAGRA\textquoteright{}s Binary Black Holes
  Form Dynamically}},}\ }\href {\doibase 10.3847/1538-4357/ac9798} {\bibfield
  {journal} {\bibinfo  {journal} {Astrophys. J.}\ }\textbf {\bibinfo {volume}
  {940}},\ \bibinfo {pages} {171} (\bibinfo {year} {2022})},\ \Eprint
  {http://arxiv.org/abs/2206.14695} {arXiv:2206.14695 [astro-ph.HE]}
  \BibitemShut {NoStop}%
\bibitem [{\citenamefont {Samsing}\ \emph {et~al.}(2022)\citenamefont
  {Samsing}, \citenamefont {Bartos}, \citenamefont {D'Orazio}, \citenamefont
  {Haiman}, \citenamefont {Kocsis}, \citenamefont {Leigh}, \citenamefont {Liu},
  \citenamefont {Pessah},\ and\ \citenamefont {Tagawa}}]{Samsing:2020tda}%
  \BibitemOpen
  \bibfield  {author} {\bibinfo {author} {\bibfnamefont {J.}~\bibnamefont
  {Samsing}}, \bibinfo {author} {\bibfnamefont {I.}~\bibnamefont {Bartos}},
  \bibinfo {author} {\bibfnamefont {D.~J.}\ \bibnamefont {D'Orazio}}, \bibinfo
  {author} {\bibfnamefont {Z.}~\bibnamefont {Haiman}}, \bibinfo {author}
  {\bibfnamefont {B.}~\bibnamefont {Kocsis}}, \bibinfo {author} {\bibfnamefont
  {N.~W.~C.}\ \bibnamefont {Leigh}}, \bibinfo {author} {\bibfnamefont
  {B.}~\bibnamefont {Liu}}, \bibinfo {author} {\bibfnamefont {M.~E.}\
  \bibnamefont {Pessah}}, \ and\ \bibinfo {author} {\bibfnamefont
  {H.}~\bibnamefont {Tagawa}},\ }\bibfield  {title} {\enquote {\bibinfo {title}
  {{AGN as potential factories for eccentric black hole mergers}},}\ }\href
  {\doibase 10.1038/s41586-021-04333-1} {\bibfield  {journal} {\bibinfo
  {journal} {Nature}\ }\textbf {\bibinfo {volume} {603}},\ \bibinfo {pages}
  {237--240} (\bibinfo {year} {2022})},\ \Eprint
  {http://arxiv.org/abs/2010.09765} {arXiv:2010.09765 [astro-ph.HE]}
  \BibitemShut {NoStop}%
\bibitem [{\citenamefont {Tagawa}\ \emph {et~al.}(2021)\citenamefont {Tagawa},
  \citenamefont {Kocsis}, \citenamefont {Haiman}, \citenamefont {Bartos},
  \citenamefont {Omukai},\ and\ \citenamefont {Samsing}}]{Tagawa:2020jnc}%
  \BibitemOpen
  \bibfield  {author} {\bibinfo {author} {\bibfnamefont {Hiromichi}\
  \bibnamefont {Tagawa}}, \bibinfo {author} {\bibfnamefont {Bence}\
  \bibnamefont {Kocsis}}, \bibinfo {author} {\bibfnamefont {Zoltan}\
  \bibnamefont {Haiman}}, \bibinfo {author} {\bibfnamefont {Imre}\ \bibnamefont
  {Bartos}}, \bibinfo {author} {\bibfnamefont {Kazuyuki}\ \bibnamefont
  {Omukai}}, \ and\ \bibinfo {author} {\bibfnamefont {Johan}\ \bibnamefont
  {Samsing}},\ }\bibfield  {title} {\enquote {\bibinfo {title} {{Eccentric
  Black Hole Mergers in Active Galactic Nuclei}},}\ }\href {\doibase
  10.3847/2041-8213/abd4d3} {\bibfield  {journal} {\bibinfo  {journal}
  {Astrophys. J. Lett.}\ }\textbf {\bibinfo {volume} {907}},\ \bibinfo {pages}
  {L20} (\bibinfo {year} {2021})},\ \Eprint {http://arxiv.org/abs/2010.10526}
  {arXiv:2010.10526 [astro-ph.HE]} \BibitemShut {NoStop}%
\bibitem [{\citenamefont {Kozai}(1962)}]{Kozai:1962zz}%
  \BibitemOpen
  \bibfield  {author} {\bibinfo {author} {\bibfnamefont {Yoshihide}\
  \bibnamefont {Kozai}},\ }\bibfield  {title} {\enquote {\bibinfo {title}
  {{Secular perturbations of asteroids with high inclination and
  eccentricity}},}\ }\href {\doibase 10.1086/108790} {\bibfield  {journal}
  {\bibinfo  {journal} {Astron. J.}\ }\textbf {\bibinfo {volume} {67}},\
  \bibinfo {pages} {591--598} (\bibinfo {year} {1962})}\BibitemShut {NoStop}%
\bibitem [{\citenamefont {Lidov}(1962)}]{Lidov:1962wjn}%
  \BibitemOpen
  \bibfield  {author} {\bibinfo {author} {\bibfnamefont {M.~L.}\ \bibnamefont
  {Lidov}},\ }\bibfield  {title} {\enquote {\bibinfo {title} {{The evolution of
  orbits of artificial satellites of planets under the action of gravitational
  perturbations of external bodies}},}\ }\href {\doibase
  10.1016/0032-0633(62)90129-0} {\bibfield  {journal} {\bibinfo  {journal}
  {Planet. Space Sci.}\ }\textbf {\bibinfo {volume} {9}},\ \bibinfo {pages}
  {719--759} (\bibinfo {year} {1962})}\BibitemShut {NoStop}%
\bibitem [{\citenamefont {{Naoz}}(2016)}]{Naoz:2016tri}%
  \BibitemOpen
  \bibfield  {author} {\bibinfo {author} {\bibfnamefont {Smadar}\ \bibnamefont
  {{Naoz}}},\ }\bibfield  {title} {\enquote {\bibinfo {title} {{The Eccentric
  Kozai-Lidov Effect and Its Applications}},}\ }\href {\doibase
  10.1146/annurev-astro-081915-023315} {\bibfield  {journal} {\bibinfo
  {journal} {Annual Review of Astronomy and Astrophysics}\ }\textbf {\bibinfo
  {volume} {54}},\ \bibinfo {pages} {441--489} (\bibinfo {year} {2016})},\
  \Eprint {http://arxiv.org/abs/1601.07175} {arXiv:1601.07175 [astro-ph.EP]}
  \BibitemShut {NoStop}%
\bibitem [{\citenamefont {Antonini}\ \emph {et~al.}(2017)\citenamefont
  {Antonini}, \citenamefont {Toonen},\ and\ \citenamefont
  {Hamers}}]{Antonini:2017ash}%
  \BibitemOpen
  \bibfield  {author} {\bibinfo {author} {\bibfnamefont {Fabio}\ \bibnamefont
  {Antonini}}, \bibinfo {author} {\bibfnamefont {Silvia}\ \bibnamefont
  {Toonen}}, \ and\ \bibinfo {author} {\bibfnamefont {Adrian~S.}\ \bibnamefont
  {Hamers}},\ }\bibfield  {title} {\enquote {\bibinfo {title} {{Binary black
  hole mergers from field triples: properties, rates and the impact of stellar
  evolution}},}\ }\href {\doibase 10.3847/1538-4357/aa6f5e} {\bibfield
  {journal} {\bibinfo  {journal} {Astrophys. J.}\ }\textbf {\bibinfo {volume}
  {841}},\ \bibinfo {pages} {77} (\bibinfo {year} {2017})},\ \Eprint
  {http://arxiv.org/abs/1703.06614} {arXiv:1703.06614 [astro-ph.GA]}
  \BibitemShut {NoStop}%
\bibitem [{\citenamefont {Randall}\ and\ \citenamefont
  {Xianyu}(2018)}]{Randall:2017jop}%
  \BibitemOpen
  \bibfield  {author} {\bibinfo {author} {\bibfnamefont {Lisa}\ \bibnamefont
  {Randall}}\ and\ \bibinfo {author} {\bibfnamefont {Zhong-Zhi}\ \bibnamefont
  {Xianyu}},\ }\bibfield  {title} {\enquote {\bibinfo {title} {{Induced
  Ellipticity for Inspiraling Binary Systems}},}\ }\href {\doibase
  10.3847/1538-4357/aaa1a2} {\bibfield  {journal} {\bibinfo  {journal}
  {Astrophys. J.}\ }\textbf {\bibinfo {volume} {853}},\ \bibinfo {pages} {93}
  (\bibinfo {year} {2018})},\ \Eprint {http://arxiv.org/abs/1708.08569}
  {arXiv:1708.08569 [gr-qc]} \BibitemShut {NoStop}%
\bibitem [{\citenamefont {Yu}\ \emph {et~al.}(2020)\citenamefont {Yu},
  \citenamefont {Ma}, \citenamefont {Giesler},\ and\ \citenamefont
  {Chen}}]{Yu:2020iqj}%
  \BibitemOpen
  \bibfield  {author} {\bibinfo {author} {\bibfnamefont {Hang}\ \bibnamefont
  {Yu}}, \bibinfo {author} {\bibfnamefont {Sizheng}\ \bibnamefont {Ma}},
  \bibinfo {author} {\bibfnamefont {Matthew}\ \bibnamefont {Giesler}}, \ and\
  \bibinfo {author} {\bibfnamefont {Yanbei}\ \bibnamefont {Chen}},\ }\bibfield
  {title} {\enquote {\bibinfo {title} {{Spin and Eccentricity Evolution in
  Triple Systems: from the Lidov-Kozai Interaction to the Final Merger of the
  Inner Binary}},}\ }\href {\doibase 10.1103/PhysRevD.102.123009} {\bibfield
  {journal} {\bibinfo  {journal} {Phys. Rev. D}\ }\textbf {\bibinfo {volume}
  {102}},\ \bibinfo {pages} {123009} (\bibinfo {year} {2020})},\ \Eprint
  {http://arxiv.org/abs/2007.12978} {arXiv:2007.12978 [gr-qc]} \BibitemShut
  {NoStop}%
\bibitem [{\citenamefont {Bartos}\ \emph {et~al.}(2023)\citenamefont {Bartos},
  \citenamefont {Rosswog}, \citenamefont {Gayathri}, \citenamefont {Miller},
  \citenamefont {Veske},\ and\ \citenamefont {Marka}}]{Bartos:2023lfu}%
  \BibitemOpen
  \bibfield  {author} {\bibinfo {author} {\bibfnamefont {I.}~\bibnamefont
  {Bartos}}, \bibinfo {author} {\bibfnamefont {S.}~\bibnamefont {Rosswog}},
  \bibinfo {author} {\bibfnamefont {V.}~\bibnamefont {Gayathri}}, \bibinfo
  {author} {\bibfnamefont {M.~C.}\ \bibnamefont {Miller}}, \bibinfo {author}
  {\bibfnamefont {D.}~\bibnamefont {Veske}}, \ and\ \bibinfo {author}
  {\bibfnamefont {S.}~\bibnamefont {Marka}},\ }\bibfield  {title} {\enquote
  {\bibinfo {title} {{Hierarchical Triples as Early Sources of $r$-process
  Elements}},}\ }\href@noop {} {\  (\bibinfo {year} {2023})},\ \Eprint
  {http://arxiv.org/abs/2302.10350} {arXiv:2302.10350 [astro-ph.HE]}
  \BibitemShut {NoStop}%
\bibitem [{\citenamefont {Saini}\ \emph {et~al.}(2022)\citenamefont {Saini},
  \citenamefont {Favata},\ and\ \citenamefont {Arun}}]{Saini:2022igm}%
  \BibitemOpen
  \bibfield  {author} {\bibinfo {author} {\bibfnamefont {Pankaj}\ \bibnamefont
  {Saini}}, \bibinfo {author} {\bibfnamefont {Marc}\ \bibnamefont {Favata}}, \
  and\ \bibinfo {author} {\bibfnamefont {K.~G.}\ \bibnamefont {Arun}},\
  }\bibfield  {title} {\enquote {\bibinfo {title} {{Systematic bias on
  parametrized tests of general relativity due to neglect of orbital
  eccentricity}},}\ }\href {\doibase 10.1103/PhysRevD.106.084031} {\bibfield
  {journal} {\bibinfo  {journal} {Phys. Rev. D}\ }\textbf {\bibinfo {volume}
  {106}},\ \bibinfo {pages} {084031} (\bibinfo {year} {2022})},\ \Eprint
  {http://arxiv.org/abs/2203.04634} {arXiv:2203.04634 [gr-qc]} \BibitemShut
  {NoStop}%
\bibitem [{\citenamefont {Saini}\ \emph {et~al.}(2023)\citenamefont {Saini},
  \citenamefont {Bhat}, \citenamefont {Favata},\ and\ \citenamefont
  {Arun}}]{Saini:2023rto}%
  \BibitemOpen
  \bibfield  {author} {\bibinfo {author} {\bibfnamefont {Pankaj}\ \bibnamefont
  {Saini}}, \bibinfo {author} {\bibfnamefont {Sajad~A.}\ \bibnamefont {Bhat}},
  \bibinfo {author} {\bibfnamefont {Marc}\ \bibnamefont {Favata}}, \ and\
  \bibinfo {author} {\bibfnamefont {K.~G.}\ \bibnamefont {Arun}},\ }\bibfield
  {title} {\enquote {\bibinfo {title} {{Eccentricity-induced systematic error
  on parametrized tests of general relativity: hierarchical Bayesian inference
  applied to a binary black hole population}},}\ }\href@noop {} {\  (\bibinfo
  {year} {2023})},\ \Eprint {http://arxiv.org/abs/2311.08033} {arXiv:2311.08033
  [gr-qc]} \BibitemShut {NoStop}%
\bibitem [{\citenamefont {Favata}(2014)}]{Favata:2013rwa}%
  \BibitemOpen
  \bibfield  {author} {\bibinfo {author} {\bibfnamefont {Marc}\ \bibnamefont
  {Favata}},\ }\bibfield  {title} {\enquote {\bibinfo {title} {{Systematic
  parameter errors in inspiraling neutron star binaries}},}\ }\href {\doibase
  10.1103/PhysRevLett.112.101101} {\bibfield  {journal} {\bibinfo  {journal}
  {Phys. Rev. Lett.}\ }\textbf {\bibinfo {volume} {112}},\ \bibinfo {pages}
  {101101} (\bibinfo {year} {2014})},\ \Eprint {http://arxiv.org/abs/1310.8288}
  {arXiv:1310.8288 [gr-qc]} \BibitemShut {NoStop}%
\bibitem [{\citenamefont {Favata}\ \emph {et~al.}(2022)\citenamefont {Favata},
  \citenamefont {Kim}, \citenamefont {Arun}, \citenamefont {Kim},\ and\
  \citenamefont {Lee}}]{Favata:2021vhw}%
  \BibitemOpen
  \bibfield  {author} {\bibinfo {author} {\bibfnamefont {Marc}\ \bibnamefont
  {Favata}}, \bibinfo {author} {\bibfnamefont {Chunglee}\ \bibnamefont {Kim}},
  \bibinfo {author} {\bibfnamefont {K.~G.}\ \bibnamefont {Arun}}, \bibinfo
  {author} {\bibfnamefont {JeongCho}\ \bibnamefont {Kim}}, \ and\ \bibinfo
  {author} {\bibfnamefont {Hyung~Won}\ \bibnamefont {Lee}},\ }\bibfield
  {title} {\enquote {\bibinfo {title} {{Constraining the orbital eccentricity
  of inspiralling compact binary systems with Advanced LIGO}},}\ }\href
  {\doibase 10.1103/PhysRevD.105.023003} {\bibfield  {journal} {\bibinfo
  {journal} {Phys. Rev. D}\ }\textbf {\bibinfo {volume} {105}},\ \bibinfo
  {pages} {023003} (\bibinfo {year} {2022})},\ \Eprint
  {http://arxiv.org/abs/2108.05861} {arXiv:2108.05861 [gr-qc]} \BibitemShut
  {NoStop}%
\bibitem [{\citenamefont {O'Shea}\ and\ \citenamefont
  {Kumar}(2023)}]{OShea:2021faf}%
  \BibitemOpen
  \bibfield  {author} {\bibinfo {author} {\bibfnamefont {Eamonn}\ \bibnamefont
  {O'Shea}}\ and\ \bibinfo {author} {\bibfnamefont {Prayush}\ \bibnamefont
  {Kumar}},\ }\bibfield  {title} {\enquote {\bibinfo {title} {{Correlations in
  gravitational-wave reconstructions from eccentric binaries: A case study with
  GW151226 and GW170608}},}\ }\href {\doibase 10.1103/PhysRevD.108.104018}
  {\bibfield  {journal} {\bibinfo  {journal} {Phys. Rev. D}\ }\textbf {\bibinfo
  {volume} {108}},\ \bibinfo {pages} {104018} (\bibinfo {year} {2023})},\
  \Eprint {http://arxiv.org/abs/2107.07981} {arXiv:2107.07981 [astro-ph.HE]}
  \BibitemShut {NoStop}%
\bibitem [{\citenamefont {Abbott}\ \emph {et~al.}(2018)\citenamefont {Abbott}
  \emph {et~al.}}]{KAGRA:2013rdx}%
  \BibitemOpen
  \bibfield  {author} {\bibinfo {author} {\bibfnamefont {B.~P.}\ \bibnamefont
  {Abbott}} \emph {et~al.} (\bibinfo {collaboration} {KAGRA, LIGO Scientific,
  Virgo, VIRGO}),\ }\bibfield  {title} {\enquote {\bibinfo {title} {{Prospects
  for observing and localizing gravitational-wave transients with Advanced
  LIGO, Advanced Virgo and KAGRA}},}\ }\href {\doibase
  10.1007/s41114-020-00026-9} {\bibfield  {journal} {\bibinfo  {journal}
  {Living Rev. Rel.}\ }\textbf {\bibinfo {volume} {21}},\ \bibinfo {pages} {3}
  (\bibinfo {year} {2018})},\ \Eprint {http://arxiv.org/abs/1304.0670}
  {arXiv:1304.0670 [gr-qc]} \BibitemShut {NoStop}%
\bibitem [{\citenamefont {Cutler}\ and\ \citenamefont
  {Flanagan}(1994)}]{Cutler:1994ys}%
  \BibitemOpen
  \bibfield  {author} {\bibinfo {author} {\bibfnamefont {Curt}\ \bibnamefont
  {Cutler}}\ and\ \bibinfo {author} {\bibfnamefont {Eanna~E.}\ \bibnamefont
  {Flanagan}},\ }\bibfield  {title} {\enquote {\bibinfo {title} {{Gravitational
  waves from merging compact binaries: How accurately can one extract the
  binary's parameters from the inspiral wave form?}}}\ }\href {\doibase
  10.1103/PhysRevD.49.2658} {\bibfield  {journal} {\bibinfo  {journal} {Phys.
  Rev. D}\ }\textbf {\bibinfo {volume} {49}},\ \bibinfo {pages} {2658--2697}
  (\bibinfo {year} {1994})},\ \Eprint {http://arxiv.org/abs/gr-qc/9402014}
  {arXiv:gr-qc/9402014} \BibitemShut {NoStop}%
\bibitem [{\citenamefont {Cutler}\ and\ \citenamefont
  {Vallisneri}(2007)}]{Cutler:2007mi}%
  \BibitemOpen
  \bibfield  {author} {\bibinfo {author} {\bibfnamefont {Curt}\ \bibnamefont
  {Cutler}}\ and\ \bibinfo {author} {\bibfnamefont {Michele}\ \bibnamefont
  {Vallisneri}},\ }\bibfield  {title} {\enquote {\bibinfo {title} {{LISA
  detections of massive black hole inspirals: Parameter extraction errors due
  to inaccurate template waveforms}},}\ }\href {\doibase
  10.1103/PhysRevD.76.104018} {\bibfield  {journal} {\bibinfo  {journal} {Phys.
  Rev. D}\ }\textbf {\bibinfo {volume} {76}},\ \bibinfo {pages} {104018}
  (\bibinfo {year} {2007})},\ \Eprint {http://arxiv.org/abs/0707.2982}
  {arXiv:0707.2982 [gr-qc]} \BibitemShut {NoStop}%
\bibitem [{\citenamefont {Bhat}\ \emph {et~al.}(2023)\citenamefont {Bhat},
  \citenamefont {Saini}, \citenamefont {Favata},\ and\ \citenamefont
  {Arun}}]{Bhat:2022amc}%
  \BibitemOpen
  \bibfield  {author} {\bibinfo {author} {\bibfnamefont {Sajad~A.}\
  \bibnamefont {Bhat}}, \bibinfo {author} {\bibfnamefont {Pankaj}\ \bibnamefont
  {Saini}}, \bibinfo {author} {\bibfnamefont {Marc}\ \bibnamefont {Favata}}, \
  and\ \bibinfo {author} {\bibfnamefont {K.~G.}\ \bibnamefont {Arun}},\
  }\bibfield  {title} {\enquote {\bibinfo {title} {{Systematic bias on the
  inspiral-merger-ringdown consistency test due to neglect of orbital
  eccentricity}},}\ }\href {\doibase 10.1103/PhysRevD.107.024009} {\bibfield
  {journal} {\bibinfo  {journal} {Phys. Rev. D}\ }\textbf {\bibinfo {volume}
  {107}},\ \bibinfo {pages} {024009} (\bibinfo {year} {2023})},\ \Eprint
  {http://arxiv.org/abs/2207.13761} {arXiv:2207.13761 [gr-qc]} \BibitemShut
  {NoStop}%
\bibitem [{\citenamefont {Narayan}\ \emph {et~al.}(2023)\citenamefont
  {Narayan}, \citenamefont {Johnson-McDaniel},\ and\ \citenamefont
  {Gupta}}]{Narayan:2023vhm}%
  \BibitemOpen
  \bibfield  {author} {\bibinfo {author} {\bibfnamefont {Purnima}\ \bibnamefont
  {Narayan}}, \bibinfo {author} {\bibfnamefont {Nathan~K.}\ \bibnamefont
  {Johnson-McDaniel}}, \ and\ \bibinfo {author} {\bibfnamefont {Anuradha}\
  \bibnamefont {Gupta}},\ }\bibfield  {title} {\enquote {\bibinfo {title}
  {{Effect of ignoring eccentricity in testing general relativity with
  gravitational waves}},}\ }\href {\doibase 10.1103/PhysRevD.108.064003}
  {\bibfield  {journal} {\bibinfo  {journal} {Phys. Rev. D}\ }\textbf {\bibinfo
  {volume} {108}},\ \bibinfo {pages} {064003} (\bibinfo {year} {2023})},\
  \Eprint {http://arxiv.org/abs/2306.04068} {arXiv:2306.04068 [gr-qc]}
  \BibitemShut {NoStop}%
\bibitem [{\citenamefont {Nagar}\ \emph {et~al.}(2018)\citenamefont {Nagar}
  \emph {et~al.}}]{Nagar:2018zoe}%
  \BibitemOpen
  \bibfield  {author} {\bibinfo {author} {\bibfnamefont {Alessandro}\
  \bibnamefont {Nagar}} \emph {et~al.},\ }\bibfield  {title} {\enquote
  {\bibinfo {title} {{Time-domain effective-one-body gravitational waveforms
  for coalescing compact binaries with nonprecessing spins, tides and self-spin
  effects}},}\ }\href {\doibase 10.1103/PhysRevD.98.104052} {\bibfield
  {journal} {\bibinfo  {journal} {Phys. Rev. D}\ }\textbf {\bibinfo {volume}
  {98}},\ \bibinfo {pages} {104052} (\bibinfo {year} {2018})},\ \Eprint
  {http://arxiv.org/abs/1806.01772} {arXiv:1806.01772 [gr-qc]} \BibitemShut
  {NoStop}%
\bibitem [{\citenamefont {Nagar}\ \emph
  {et~al.}(2021{\natexlab{a}})\citenamefont {Nagar}, \citenamefont {Bonino},\
  and\ \citenamefont {Rettegno}}]{Nagar:2021gss}%
  \BibitemOpen
  \bibfield  {author} {\bibinfo {author} {\bibfnamefont {Alessandro}\
  \bibnamefont {Nagar}}, \bibinfo {author} {\bibfnamefont {Alice}\ \bibnamefont
  {Bonino}}, \ and\ \bibinfo {author} {\bibfnamefont {Piero}\ \bibnamefont
  {Rettegno}},\ }\bibfield  {title} {\enquote {\bibinfo {title} {{Effective
  one-body multipolar waveform model for spin-aligned, quasicircular,
  eccentric, hyperbolic black hole binaries}},}\ }\href {\doibase
  10.1103/PhysRevD.103.104021} {\bibfield  {journal} {\bibinfo  {journal}
  {Phys. Rev. D}\ }\textbf {\bibinfo {volume} {103}},\ \bibinfo {pages}
  {104021} (\bibinfo {year} {2021}{\natexlab{a}})},\ \Eprint
  {http://arxiv.org/abs/2101.08624} {arXiv:2101.08624 [gr-qc]} \BibitemShut
  {NoStop}%
\bibitem [{\citenamefont {Shaikh}\ \emph {et~al.}(2023)\citenamefont {Shaikh},
  \citenamefont {Varma}, \citenamefont {Pfeiffer}, \citenamefont
  {Ramos-Buades},\ and\ \citenamefont {van~de Meent}}]{Shaikh:2023ypz}%
  \BibitemOpen
  \bibfield  {author} {\bibinfo {author} {\bibfnamefont {Md~Arif}\ \bibnamefont
  {Shaikh}}, \bibinfo {author} {\bibfnamefont {Vijay}\ \bibnamefont {Varma}},
  \bibinfo {author} {\bibfnamefont {Harald~P.}\ \bibnamefont {Pfeiffer}},
  \bibinfo {author} {\bibfnamefont {Antoni}\ \bibnamefont {Ramos-Buades}}, \
  and\ \bibinfo {author} {\bibfnamefont {Maarten}\ \bibnamefont {van~de
  Meent}},\ }\bibfield  {title} {\enquote {\bibinfo {title} {{Defining
  eccentricity for gravitational wave astronomy}},}\ }\href {\doibase
  10.1103/PhysRevD.108.104007} {\bibfield  {journal} {\bibinfo  {journal}
  {Phys. Rev. D}\ }\textbf {\bibinfo {volume} {108}},\ \bibinfo {pages}
  {104007} (\bibinfo {year} {2023})},\ \bibinfo {note}
  {{\href{https://pypi.org/project/gw_eccentricity}{pypi.org/project/gw\_eccentricity}}},\
  \Eprint {http://arxiv.org/abs/2302.11257} {arXiv:2302.11257 [gr-qc]}
  \BibitemShut {NoStop}%
\bibitem [{\citenamefont {Clarke}\ \emph {et~al.}(2022)\citenamefont {Clarke},
  \citenamefont {Romero-Shaw}, \citenamefont {Lasky},\ and\ \citenamefont
  {Thrane}}]{Clarke:2022fma}%
  \BibitemOpen
  \bibfield  {author} {\bibinfo {author} {\bibfnamefont {Teagan~A.}\
  \bibnamefont {Clarke}}, \bibinfo {author} {\bibfnamefont {Isobel~M.}\
  \bibnamefont {Romero-Shaw}}, \bibinfo {author} {\bibfnamefont {Paul~D.}\
  \bibnamefont {Lasky}}, \ and\ \bibinfo {author} {\bibfnamefont {Eric}\
  \bibnamefont {Thrane}},\ }\bibfield  {title} {\enquote {\bibinfo {title}
  {{Gravitational-wave inference for eccentric binaries: the argument of
  periapsis}},}\ }\href {\doibase 10.1093/mnras/stac2965} {\bibfield  {journal}
  {\bibinfo  {journal} {Mon. Not. Roy. Astron. Soc.}\ }\textbf {\bibinfo
  {volume} {517}},\ \bibinfo {pages} {3778--3784} (\bibinfo {year} {2022})},\
  \Eprint {http://arxiv.org/abs/2206.14006} {arXiv:2206.14006 [gr-qc]}
  \BibitemShut {NoStop}%
\bibitem [{\citenamefont {Hofmann}\ \emph {et~al.}(2016)\citenamefont
  {Hofmann}, \citenamefont {Barausse},\ and\ \citenamefont
  {Rezzolla}}]{Hofmann:2016yih}%
  \BibitemOpen
  \bibfield  {author} {\bibinfo {author} {\bibfnamefont {Fabian}\ \bibnamefont
  {Hofmann}}, \bibinfo {author} {\bibfnamefont {Enrico}\ \bibnamefont
  {Barausse}}, \ and\ \bibinfo {author} {\bibfnamefont {Luciano}\ \bibnamefont
  {Rezzolla}},\ }\bibfield  {title} {\enquote {\bibinfo {title} {{The final
  spin from binary black holes in quasi-circular orbits}},}\ }\href {\doibase
  10.3847/2041-8205/825/2/L19} {\bibfield  {journal} {\bibinfo  {journal}
  {Astrophys. J. Lett.}\ }\textbf {\bibinfo {volume} {825}},\ \bibinfo {pages}
  {L19} (\bibinfo {year} {2016})},\ \Eprint {http://arxiv.org/abs/1605.01938}
  {arXiv:1605.01938 [gr-qc]} \BibitemShut {NoStop}%
\bibitem [{\citenamefont {Healy}\ and\ \citenamefont
  {Lousto}(2017)}]{Healy:2016lce}%
  \BibitemOpen
  \bibfield  {author} {\bibinfo {author} {\bibfnamefont {James}\ \bibnamefont
  {Healy}}\ and\ \bibinfo {author} {\bibfnamefont {Carlos~O.}\ \bibnamefont
  {Lousto}},\ }\bibfield  {title} {\enquote {\bibinfo {title} {{Remnant of
  binary black-hole mergers: New simulations and peak luminosity studies}},}\
  }\href {\doibase 10.1103/PhysRevD.95.024037} {\bibfield  {journal} {\bibinfo
  {journal} {Phys. Rev. D}\ }\textbf {\bibinfo {volume} {95}},\ \bibinfo
  {pages} {024037} (\bibinfo {year} {2017})},\ \Eprint
  {http://arxiv.org/abs/1610.09713} {arXiv:1610.09713 [gr-qc]} \BibitemShut
  {NoStop}%
\bibitem [{\citenamefont {Jim\'enez-Forteza}\ \emph {et~al.}(2017)\citenamefont
  {Jim\'enez-Forteza}, \citenamefont {Keitel}, \citenamefont {Husa},
  \citenamefont {Hannam}, \citenamefont {Khan},\ and\ \citenamefont
  {P\"urrer}}]{Jimenez-Forteza:2016oae}%
  \BibitemOpen
  \bibfield  {author} {\bibinfo {author} {\bibfnamefont {Xisco}\ \bibnamefont
  {Jim\'enez-Forteza}}, \bibinfo {author} {\bibfnamefont {David}\ \bibnamefont
  {Keitel}}, \bibinfo {author} {\bibfnamefont {Sascha}\ \bibnamefont {Husa}},
  \bibinfo {author} {\bibfnamefont {Mark}\ \bibnamefont {Hannam}}, \bibinfo
  {author} {\bibfnamefont {Sebastian}\ \bibnamefont {Khan}}, \ and\ \bibinfo
  {author} {\bibfnamefont {Michael}\ \bibnamefont {P\"urrer}},\ }\bibfield
  {title} {\enquote {\bibinfo {title} {{Hierarchical data-driven approach to
  fitting numerical relativity data for nonprecessing binary black holes with
  an application to final spin and radiated energy}},}\ }\href {\doibase
  10.1103/PhysRevD.95.064024} {\bibfield  {journal} {\bibinfo  {journal} {Phys.
  Rev. D}\ }\textbf {\bibinfo {volume} {95}},\ \bibinfo {pages} {064024}
  (\bibinfo {year} {2017})},\ \Eprint {http://arxiv.org/abs/1611.00332}
  {arXiv:1611.00332 [gr-qc]} \BibitemShut {NoStop}%
\bibitem [{\citenamefont {Mora}\ and\ \citenamefont
  {Will}(2002)}]{Mora:2002gf}%
  \BibitemOpen
  \bibfield  {author} {\bibinfo {author} {\bibfnamefont {Thierry}\ \bibnamefont
  {Mora}}\ and\ \bibinfo {author} {\bibfnamefont {Clifford~M.}\ \bibnamefont
  {Will}},\ }\bibfield  {title} {\enquote {\bibinfo {title} {{Numerically
  generated quasiequilibrium orbits of black holes: Circular or eccentric?}}}\
  }\href {\doibase 10.1103/PhysRevD.66.101501} {\bibfield  {journal} {\bibinfo
  {journal} {Phys. Rev. D}\ }\textbf {\bibinfo {volume} {66}},\ \bibinfo
  {pages} {101501} (\bibinfo {year} {2002})},\ \Eprint
  {http://arxiv.org/abs/gr-qc/0208089} {arXiv:gr-qc/0208089} \BibitemShut
  {NoStop}%
\bibitem [{\citenamefont {Ramos-Buades}\ \emph {et~al.}(2020)\citenamefont
  {Ramos-Buades}, \citenamefont {Husa}, \citenamefont {Pratten}, \citenamefont
  {Estell\'es}, \citenamefont {Garc\'\i{}a-Quir\'os}, \citenamefont
  {Mateu-Lucena}, \citenamefont {Colleoni},\ and\ \citenamefont
  {Jaume}}]{Ramos-Buades:2019uvh}%
  \BibitemOpen
  \bibfield  {author} {\bibinfo {author} {\bibfnamefont {Antoni}\ \bibnamefont
  {Ramos-Buades}}, \bibinfo {author} {\bibfnamefont {Sascha}\ \bibnamefont
  {Husa}}, \bibinfo {author} {\bibfnamefont {Geraint}\ \bibnamefont {Pratten}},
  \bibinfo {author} {\bibfnamefont {H\'ector}\ \bibnamefont {Estell\'es}},
  \bibinfo {author} {\bibfnamefont {Cecilio}\ \bibnamefont
  {Garc\'\i{}a-Quir\'os}}, \bibinfo {author} {\bibfnamefont {Maite}\
  \bibnamefont {Mateu-Lucena}}, \bibinfo {author} {\bibfnamefont {Marta}\
  \bibnamefont {Colleoni}}, \ and\ \bibinfo {author} {\bibfnamefont {Rafel}\
  \bibnamefont {Jaume}},\ }\bibfield  {title} {\enquote {\bibinfo {title}
  {{First survey of spinning eccentric black hole mergers: Numerical relativity
  simulations, hybrid waveforms, and parameter estimation}},}\ }\href {\doibase
  10.1103/PhysRevD.101.083015} {\bibfield  {journal} {\bibinfo  {journal}
  {Phys. Rev. D}\ }\textbf {\bibinfo {volume} {101}},\ \bibinfo {pages}
  {083015} (\bibinfo {year} {2020})},\ \Eprint
  {http://arxiv.org/abs/1909.11011} {arXiv:1909.11011 [gr-qc]} \BibitemShut
  {NoStop}%
\bibitem [{\citenamefont {Islam}\ \emph {et~al.}(2021)\citenamefont {Islam},
  \citenamefont {Varma}, \citenamefont {Lodman}, \citenamefont {Field},
  \citenamefont {Khanna}, \citenamefont {Scheel}, \citenamefont {Pfeiffer},
  \citenamefont {Gerosa},\ and\ \citenamefont {Kidder}}]{Islam:2021mha}%
  \BibitemOpen
  \bibfield  {author} {\bibinfo {author} {\bibfnamefont {Tousif}\ \bibnamefont
  {Islam}}, \bibinfo {author} {\bibfnamefont {Vijay}\ \bibnamefont {Varma}},
  \bibinfo {author} {\bibfnamefont {Jackie}\ \bibnamefont {Lodman}}, \bibinfo
  {author} {\bibfnamefont {Scott~E.}\ \bibnamefont {Field}}, \bibinfo {author}
  {\bibfnamefont {Gaurav}\ \bibnamefont {Khanna}}, \bibinfo {author}
  {\bibfnamefont {Mark~A.}\ \bibnamefont {Scheel}}, \bibinfo {author}
  {\bibfnamefont {Harald~P.}\ \bibnamefont {Pfeiffer}}, \bibinfo {author}
  {\bibfnamefont {Davide}\ \bibnamefont {Gerosa}}, \ and\ \bibinfo {author}
  {\bibfnamefont {Lawrence~E.}\ \bibnamefont {Kidder}},\ }\bibfield  {title}
  {\enquote {\bibinfo {title} {{Eccentric binary black hole surrogate models
  for the gravitational waveform and remnant properties: comparable mass,
  nonspinning case}},}\ }\href {\doibase 10.1103/PhysRevD.103.064022}
  {\bibfield  {journal} {\bibinfo  {journal} {Phys. Rev. D}\ }\textbf {\bibinfo
  {volume} {103}},\ \bibinfo {pages} {064022} (\bibinfo {year} {2021})},\
  \Eprint {http://arxiv.org/abs/2101.11798} {arXiv:2101.11798 [gr-qc]}
  \BibitemShut {NoStop}%
\bibitem [{\citenamefont {Ramos-Buades}\ \emph
  {et~al.}(2022{\natexlab{a}})\citenamefont {Ramos-Buades}, \citenamefont
  {Buonanno}, \citenamefont {Khalil},\ and\ \citenamefont
  {Ossokine}}]{Ramos-Buades:2021adz}%
  \BibitemOpen
  \bibfield  {author} {\bibinfo {author} {\bibfnamefont {Antoni}\ \bibnamefont
  {Ramos-Buades}}, \bibinfo {author} {\bibfnamefont {Alessandra}\ \bibnamefont
  {Buonanno}}, \bibinfo {author} {\bibfnamefont {Mohammed}\ \bibnamefont
  {Khalil}}, \ and\ \bibinfo {author} {\bibfnamefont {Serguei}\ \bibnamefont
  {Ossokine}},\ }\bibfield  {title} {\enquote {\bibinfo {title}
  {{Effective-one-body multipolar waveforms for eccentric binary black holes
  with nonprecessing spins}},}\ }\href {\doibase 10.1103/PhysRevD.105.044035}
  {\bibfield  {journal} {\bibinfo  {journal} {Phys. Rev. D}\ }\textbf {\bibinfo
  {volume} {105}},\ \bibinfo {pages} {044035} (\bibinfo {year}
  {2022}{\natexlab{a}})},\ \Eprint {http://arxiv.org/abs/2112.06952}
  {arXiv:2112.06952 [gr-qc]} \BibitemShut {NoStop}%
\bibitem [{\citenamefont {Bonino}\ \emph {et~al.}(2023)\citenamefont {Bonino},
  \citenamefont {Gamba}, \citenamefont {Schmidt}, \citenamefont {Nagar},
  \citenamefont {Pratten}, \citenamefont {Breschi}, \citenamefont {Rettegno},\
  and\ \citenamefont {Bernuzzi}}]{Bonino:2022hkj}%
  \BibitemOpen
  \bibfield  {author} {\bibinfo {author} {\bibfnamefont {Alice}\ \bibnamefont
  {Bonino}}, \bibinfo {author} {\bibfnamefont {Rossella}\ \bibnamefont
  {Gamba}}, \bibinfo {author} {\bibfnamefont {Patricia}\ \bibnamefont
  {Schmidt}}, \bibinfo {author} {\bibfnamefont {Alessandro}\ \bibnamefont
  {Nagar}}, \bibinfo {author} {\bibfnamefont {Geraint}\ \bibnamefont
  {Pratten}}, \bibinfo {author} {\bibfnamefont {Matteo}\ \bibnamefont
  {Breschi}}, \bibinfo {author} {\bibfnamefont {Piero}\ \bibnamefont
  {Rettegno}}, \ and\ \bibinfo {author} {\bibfnamefont {Sebastiano}\
  \bibnamefont {Bernuzzi}},\ }\bibfield  {title} {\enquote {\bibinfo {title}
  {{Inferring eccentricity evolution from observations of coalescing binary
  black holes}},}\ }\href {\doibase 10.1103/PhysRevD.107.064024} {\bibfield
  {journal} {\bibinfo  {journal} {Phys. Rev. D}\ }\textbf {\bibinfo {volume}
  {107}},\ \bibinfo {pages} {064024} (\bibinfo {year} {2023})},\ \Eprint
  {http://arxiv.org/abs/2207.10474} {arXiv:2207.10474 [gr-qc]} \BibitemShut
  {NoStop}%
\bibitem [{\citenamefont {Ramos-Buades}\ \emph
  {et~al.}(2022{\natexlab{b}})\citenamefont {Ramos-Buades}, \citenamefont
  {van~de Meent}, \citenamefont {Pfeiffer}, \citenamefont {R\"uter},
  \citenamefont {Scheel}, \citenamefont {Boyle},\ and\ \citenamefont
  {Kidder}}]{Ramos-Buades:2022lgf}%
  \BibitemOpen
  \bibfield  {author} {\bibinfo {author} {\bibfnamefont {Antoni}\ \bibnamefont
  {Ramos-Buades}}, \bibinfo {author} {\bibfnamefont {Maarten}\ \bibnamefont
  {van~de Meent}}, \bibinfo {author} {\bibfnamefont {Harald~P.}\ \bibnamefont
  {Pfeiffer}}, \bibinfo {author} {\bibfnamefont {Hannes~R.}\ \bibnamefont
  {R\"uter}}, \bibinfo {author} {\bibfnamefont {Mark~A.}\ \bibnamefont
  {Scheel}}, \bibinfo {author} {\bibfnamefont {Michael}\ \bibnamefont {Boyle}},
  \ and\ \bibinfo {author} {\bibfnamefont {Lawrence~E.}\ \bibnamefont
  {Kidder}},\ }\bibfield  {title} {\enquote {\bibinfo {title} {{Eccentric
  binary black holes: Comparing numerical relativity and small mass-ratio
  perturbation theory}},}\ }\href {\doibase 10.1103/PhysRevD.106.124040}
  {\bibfield  {journal} {\bibinfo  {journal} {Phys. Rev. D}\ }\textbf {\bibinfo
  {volume} {106}},\ \bibinfo {pages} {124040} (\bibinfo {year}
  {2022}{\natexlab{b}})},\ \Eprint {http://arxiv.org/abs/2209.03390}
  {arXiv:2209.03390 [gr-qc]} \BibitemShut {NoStop}%
\bibitem [{\citenamefont {Ashton}\ \emph {et~al.}(2019)\citenamefont {Ashton}
  \emph {et~al.}}]{Ashton:2018jfp}%
  \BibitemOpen
  \bibfield  {author} {\bibinfo {author} {\bibfnamefont {Gregory}\ \bibnamefont
  {Ashton}} \emph {et~al.},\ }\bibfield  {title} {\enquote {\bibinfo {title}
  {{BILBY: A user-friendly Bayesian inference library for gravitational-wave
  astronomy}},}\ }\href {\doibase 10.3847/1538-4365/ab06fc} {\bibfield
  {journal} {\bibinfo  {journal} {Astrophys. J. Suppl.}\ }\textbf {\bibinfo
  {volume} {241}},\ \bibinfo {pages} {27} (\bibinfo {year} {2019})},\ \Eprint
  {http://arxiv.org/abs/1811.02042} {arXiv:1811.02042 [astro-ph.IM]}
  \BibitemShut {NoStop}%
\bibitem [{\citenamefont {Speagle}(2020)}]{Speagle:2019ivv}%
  \BibitemOpen
  \bibfield  {author} {\bibinfo {author} {\bibfnamefont {Joshua~S.}\
  \bibnamefont {Speagle}},\ }\bibfield  {title} {\enquote {\bibinfo {title}
  {{dynesty: a dynamic nested sampling package for estimating Bayesian
  posteriors and evidences}},}\ }\href {\doibase 10.1093/mnras/staa278}
  {\bibfield  {journal} {\bibinfo  {journal} {Mon. Not. Roy. Astron. Soc.}\
  }\textbf {\bibinfo {volume} {493}},\ \bibinfo {pages} {3132--3158} (\bibinfo
  {year} {2020})},\ \Eprint {http://arxiv.org/abs/1904.02180} {arXiv:1904.02180
  [astro-ph.IM]} \BibitemShut {NoStop}%
\bibitem [{\citenamefont {Nagar}\ \emph
  {et~al.}(2021{\natexlab{b}})\citenamefont {Nagar}, \citenamefont {Rettegno},
  \citenamefont {Gamba},\ and\ \citenamefont {Bernuzzi}}]{Nagar:2020xsk}%
  \BibitemOpen
  \bibfield  {author} {\bibinfo {author} {\bibfnamefont {Alessandro}\
  \bibnamefont {Nagar}}, \bibinfo {author} {\bibfnamefont {Piero}\ \bibnamefont
  {Rettegno}}, \bibinfo {author} {\bibfnamefont {Rossella}\ \bibnamefont
  {Gamba}}, \ and\ \bibinfo {author} {\bibfnamefont {Sebastiano}\ \bibnamefont
  {Bernuzzi}},\ }\bibfield  {title} {\enquote {\bibinfo {title}
  {{Effective-one-body waveforms from dynamical captures in black hole
  binaries}},}\ }\href {\doibase 10.1103/PhysRevD.103.064013} {\bibfield
  {journal} {\bibinfo  {journal} {Phys. Rev. D}\ }\textbf {\bibinfo {volume}
  {103}},\ \bibinfo {pages} {064013} (\bibinfo {year} {2021}{\natexlab{b}})},\
  \Eprint {http://arxiv.org/abs/2009.12857} {arXiv:2009.12857 [gr-qc]}
  \BibitemShut {NoStop}%
\bibitem [{\citenamefont {Chiaramello}\ and\ \citenamefont
  {Nagar}(2020)}]{Chiaramello:2020ehz}%
  \BibitemOpen
  \bibfield  {author} {\bibinfo {author} {\bibfnamefont {Danilo}\ \bibnamefont
  {Chiaramello}}\ and\ \bibinfo {author} {\bibfnamefont {Alessandro}\
  \bibnamefont {Nagar}},\ }\bibfield  {title} {\enquote {\bibinfo {title}
  {{Faithful analytical effective-one-body waveform model for spin-aligned,
  moderately eccentric, coalescing black hole binaries}},}\ }\href {\doibase
  10.1103/PhysRevD.101.101501} {\bibfield  {journal} {\bibinfo  {journal}
  {Phys. Rev. D}\ }\textbf {\bibinfo {volume} {101}},\ \bibinfo {pages}
  {101501} (\bibinfo {year} {2020})},\ \Eprint
  {http://arxiv.org/abs/2001.11736} {arXiv:2001.11736 [gr-qc]} \BibitemShut
  {NoStop}%
\bibitem [{\citenamefont {Albanesi}\ \emph {et~al.}(2021)\citenamefont
  {Albanesi}, \citenamefont {Nagar},\ and\ \citenamefont
  {Bernuzzi}}]{Albanesi:2021rby}%
  \BibitemOpen
  \bibfield  {author} {\bibinfo {author} {\bibfnamefont {Simone}\ \bibnamefont
  {Albanesi}}, \bibinfo {author} {\bibfnamefont {Alessandro}\ \bibnamefont
  {Nagar}}, \ and\ \bibinfo {author} {\bibfnamefont {Sebastiano}\ \bibnamefont
  {Bernuzzi}},\ }\bibfield  {title} {\enquote {\bibinfo {title} {{Effective
  one-body model for extreme-mass-ratio spinning binaries on eccentric
  equatorial orbits: Testing radiation reaction and waveform}},}\ }\href
  {\doibase 10.1103/PhysRevD.104.024067} {\bibfield  {journal} {\bibinfo
  {journal} {Phys. Rev. D}\ }\textbf {\bibinfo {volume} {104}},\ \bibinfo
  {pages} {024067} (\bibinfo {year} {2021})},\ \Eprint
  {http://arxiv.org/abs/2104.10559} {arXiv:2104.10559 [gr-qc]} \BibitemShut
  {NoStop}%
\bibitem [{\citenamefont {Nagar}\ and\ \citenamefont
  {Rettegno}(2021)}]{Nagar:2021xnh}%
  \BibitemOpen
  \bibfield  {author} {\bibinfo {author} {\bibfnamefont {Alessandro}\
  \bibnamefont {Nagar}}\ and\ \bibinfo {author} {\bibfnamefont {Piero}\
  \bibnamefont {Rettegno}},\ }\bibfield  {title} {\enquote {\bibinfo {title}
  {{Next generation: Impact of high-order analytical information on effective
  one body waveform models for noncircularized, spin-aligned black hole
  binaries}},}\ }\href {\doibase 10.1103/PhysRevD.104.104004} {\bibfield
  {journal} {\bibinfo  {journal} {Phys. Rev. D}\ }\textbf {\bibinfo {volume}
  {104}},\ \bibinfo {pages} {104004} (\bibinfo {year} {2021})},\ \Eprint
  {http://arxiv.org/abs/2108.02043} {arXiv:2108.02043 [gr-qc]} \BibitemShut
  {NoStop}%
\bibitem [{\citenamefont {Nagar}\ \emph {et~al.}(2024)\citenamefont {Nagar},
  \citenamefont {Gamba}, \citenamefont {Rettegno}, \citenamefont {Fantini},\
  and\ \citenamefont {Bernuzzi}}]{Nagar:2024dzj}%
  \BibitemOpen
  \bibfield  {author} {\bibinfo {author} {\bibfnamefont {Alessandro}\
  \bibnamefont {Nagar}}, \bibinfo {author} {\bibfnamefont {Rossella}\
  \bibnamefont {Gamba}}, \bibinfo {author} {\bibfnamefont {Piero}\ \bibnamefont
  {Rettegno}}, \bibinfo {author} {\bibfnamefont {Veronica}\ \bibnamefont
  {Fantini}}, \ and\ \bibinfo {author} {\bibfnamefont {Sebastiano}\
  \bibnamefont {Bernuzzi}},\ }\bibfield  {title} {\enquote {\bibinfo {title}
  {{Effective-one-body waveform model for non-circularized, planar, coalescing
  black hole binaries: the importance of radiation reaction}},}\ }\href@noop {}
  {\  (\bibinfo {year} {2024})},\ \Eprint {http://arxiv.org/abs/2404.05288}
  {arXiv:2404.05288 [gr-qc]} \BibitemShut {NoStop}%
\bibitem [{\citenamefont {Ramos-Buades}\ \emph {et~al.}(2023)\citenamefont
  {Ramos-Buades}, \citenamefont {Buonanno},\ and\ \citenamefont
  {Gair}}]{Ramos-Buades:2023yhy}%
  \BibitemOpen
  \bibfield  {author} {\bibinfo {author} {\bibfnamefont {Antoni}\ \bibnamefont
  {Ramos-Buades}}, \bibinfo {author} {\bibfnamefont {Alessandra}\ \bibnamefont
  {Buonanno}}, \ and\ \bibinfo {author} {\bibfnamefont {Jonathan}\ \bibnamefont
  {Gair}},\ }\bibfield  {title} {\enquote {\bibinfo {title} {{Bayesian
  inference of binary black holes with inspiral-merger-ringdown waveforms using
  two eccentric parameters}},}\ }\href@noop {} {\  (\bibinfo {year} {2023})},\
  \Eprint {http://arxiv.org/abs/2309.15528} {arXiv:2309.15528 [gr-qc]}
  \BibitemShut {NoStop}%
\bibitem [{\citenamefont {Cao}\ and\ \citenamefont {Han}(2017)}]{Cao:2017ndf}%
  \BibitemOpen
  \bibfield  {author} {\bibinfo {author} {\bibfnamefont {Zhoujian}\
  \bibnamefont {Cao}}\ and\ \bibinfo {author} {\bibfnamefont {Wen-Biao}\
  \bibnamefont {Han}},\ }\bibfield  {title} {\enquote {\bibinfo {title}
  {{Waveform model for an eccentric binary black hole based on the
  effective-one-body-numerical-relativity formalism}},}\ }\href {\doibase
  10.1103/PhysRevD.96.044028} {\bibfield  {journal} {\bibinfo  {journal} {Phys.
  Rev. D}\ }\textbf {\bibinfo {volume} {96}},\ \bibinfo {pages} {044028}
  (\bibinfo {year} {2017})},\ \Eprint {http://arxiv.org/abs/1708.00166}
  {arXiv:1708.00166 [gr-qc]} \BibitemShut {NoStop}%
\bibitem [{\citenamefont {Liu}\ \emph {et~al.}(2020)\citenamefont {Liu},
  \citenamefont {Cao},\ and\ \citenamefont {Shao}}]{Liu:2019jpg}%
  \BibitemOpen
  \bibfield  {author} {\bibinfo {author} {\bibfnamefont {Xiaolin}\ \bibnamefont
  {Liu}}, \bibinfo {author} {\bibfnamefont {Zhoujian}\ \bibnamefont {Cao}}, \
  and\ \bibinfo {author} {\bibfnamefont {Lijing}\ \bibnamefont {Shao}},\
  }\bibfield  {title} {\enquote {\bibinfo {title} {{Validating the
  Effective-One-Body Numerical-Relativity Waveform Models for Spin-aligned
  Binary Black Holes along Eccentric Orbits}},}\ }\href {\doibase
  10.1103/PhysRevD.101.044049} {\bibfield  {journal} {\bibinfo  {journal}
  {Phys. Rev. D}\ }\textbf {\bibinfo {volume} {101}},\ \bibinfo {pages}
  {044049} (\bibinfo {year} {2020})},\ \Eprint
  {http://arxiv.org/abs/1910.00784} {arXiv:1910.00784 [gr-qc]} \BibitemShut
  {NoStop}%
\bibitem [{\citenamefont {Tanay}\ \emph {et~al.}(2016)\citenamefont {Tanay},
  \citenamefont {Haney},\ and\ \citenamefont {Gopakumar}}]{Tanay:2016zog}%
  \BibitemOpen
  \bibfield  {author} {\bibinfo {author} {\bibfnamefont {Sashwat}\ \bibnamefont
  {Tanay}}, \bibinfo {author} {\bibfnamefont {Maria}\ \bibnamefont {Haney}}, \
  and\ \bibinfo {author} {\bibfnamefont {Achamveedu}\ \bibnamefont
  {Gopakumar}},\ }\bibfield  {title} {\enquote {\bibinfo {title} {{Frequency
  and time domain inspiral templates for comparable mass compact binaries in
  eccentric orbits}},}\ }\href {\doibase 10.1103/PhysRevD.93.064031} {\bibfield
   {journal} {\bibinfo  {journal} {Phys. Rev. D}\ }\textbf {\bibinfo {volume}
  {93}},\ \bibinfo {pages} {064031} (\bibinfo {year} {2016})},\ \Eprint
  {http://arxiv.org/abs/1602.03081} {arXiv:1602.03081 [gr-qc]} \BibitemShut
  {NoStop}%
\bibitem [{\citenamefont {Carullo}\ \emph {et~al.}(2024)\citenamefont
  {Carullo}, \citenamefont {Albanesi}, \citenamefont {Nagar}, \citenamefont
  {Gamba}, \citenamefont {Bernuzzi}, \citenamefont {Andrade},\ and\
  \citenamefont {Trenado}}]{Carullo:2023kvj}%
  \BibitemOpen
  \bibfield  {author} {\bibinfo {author} {\bibfnamefont {Gregorio}\
  \bibnamefont {Carullo}}, \bibinfo {author} {\bibfnamefont {Simone}\
  \bibnamefont {Albanesi}}, \bibinfo {author} {\bibfnamefont {Alessandro}\
  \bibnamefont {Nagar}}, \bibinfo {author} {\bibfnamefont {Rossella}\
  \bibnamefont {Gamba}}, \bibinfo {author} {\bibfnamefont {Sebastiano}\
  \bibnamefont {Bernuzzi}}, \bibinfo {author} {\bibfnamefont {Tomas}\
  \bibnamefont {Andrade}}, \ and\ \bibinfo {author} {\bibfnamefont {Juan}\
  \bibnamefont {Trenado}},\ }\bibfield  {title} {\enquote {\bibinfo {title}
  {{Unveiling the Merger Structure of Black Hole Binaries in Generic Planar
  Orbits}},}\ }\href {\doibase 10.1103/PhysRevLett.132.101401} {\bibfield
  {journal} {\bibinfo  {journal} {Phys. Rev. Lett.}\ }\textbf {\bibinfo
  {volume} {132}},\ \bibinfo {pages} {101401} (\bibinfo {year} {2024})},\
  \Eprint {http://arxiv.org/abs/2309.07228} {arXiv:2309.07228 [gr-qc]}
  \BibitemShut {NoStop}%
\bibitem [{\citenamefont {Gamba}\ \emph {et~al.}(2023)\citenamefont {Gamba},
  \citenamefont {Breschi}, \citenamefont {Carullo}, \citenamefont {Albanesi},
  \citenamefont {Rettegno}, \citenamefont {Bernuzzi},\ and\ \citenamefont
  {Nagar}}]{Gamba:2021gap}%
  \BibitemOpen
  \bibfield  {author} {\bibinfo {author} {\bibfnamefont {Rossella}\
  \bibnamefont {Gamba}}, \bibinfo {author} {\bibfnamefont {Matteo}\
  \bibnamefont {Breschi}}, \bibinfo {author} {\bibfnamefont {Gregorio}\
  \bibnamefont {Carullo}}, \bibinfo {author} {\bibfnamefont {Simone}\
  \bibnamefont {Albanesi}}, \bibinfo {author} {\bibfnamefont {Piero}\
  \bibnamefont {Rettegno}}, \bibinfo {author} {\bibfnamefont {Sebastiano}\
  \bibnamefont {Bernuzzi}}, \ and\ \bibinfo {author} {\bibfnamefont
  {Alessandro}\ \bibnamefont {Nagar}},\ }\bibfield  {title} {\enquote {\bibinfo
  {title} {{GW190521 as a dynamical capture of two nonspinning black holes}},}\
  }\href {\doibase 10.1038/s41550-022-01813-w} {\bibfield  {journal} {\bibinfo
  {journal} {Nature Astron.}\ }\textbf {\bibinfo {volume} {7}},\ \bibinfo
  {pages} {11--17} (\bibinfo {year} {2023})},\ \Eprint
  {http://arxiv.org/abs/2106.05575} {arXiv:2106.05575 [gr-qc]} \BibitemShut
  {NoStop}%
\bibitem [{\citenamefont {Bonino}\ \emph {et~al.}(2024)\citenamefont {Bonino},
  \citenamefont {Schmidt},\ and\ \citenamefont {Pratten}}]{Bonino:2024xrv}%
  \BibitemOpen
  \bibfield  {author} {\bibinfo {author} {\bibfnamefont {Alice}\ \bibnamefont
  {Bonino}}, \bibinfo {author} {\bibfnamefont {Patricia}\ \bibnamefont
  {Schmidt}}, \ and\ \bibinfo {author} {\bibfnamefont {Geraint}\ \bibnamefont
  {Pratten}},\ }\bibfield  {title} {\enquote {\bibinfo {title} {{Mapping
  eccentricity evolutions between numerical relativity and effective-one-body
  gravitational waveforms}},}\ }\href@noop {} {\  (\bibinfo {year} {2024})},\
  \Eprint {http://arxiv.org/abs/2404.18875} {arXiv:2404.18875 [gr-qc]}
  \BibitemShut {NoStop}%
\bibitem [{\citenamefont {Gupte}\ \emph {et~al.}(2024)\citenamefont {Gupte}
  \emph {et~al.}}]{Gupte:2024jfe}%
  \BibitemOpen
  \bibfield  {author} {\bibinfo {author} {\bibfnamefont {Nihar}\ \bibnamefont
  {Gupte}} \emph {et~al.},\ }\bibfield  {title} {\enquote {\bibinfo {title}
  {{Evidence for eccentricity in the population of binary black holes observed
  by LIGO-Virgo-KAGRA}},}\ }\href@noop {} {\  (\bibinfo {year} {2024})},\
  \Eprint {http://arxiv.org/abs/2404.14286} {arXiv:2404.14286 [gr-qc]}
  \BibitemShut {NoStop}%
\bibitem [{\citenamefont {Hoy}\ and\ \citenamefont
  {Raymond}(2021)}]{Hoy:2020vys}%
  \BibitemOpen
  \bibfield  {author} {\bibinfo {author} {\bibfnamefont {Charlie}\ \bibnamefont
  {Hoy}}\ and\ \bibinfo {author} {\bibfnamefont {Vivien}\ \bibnamefont
  {Raymond}},\ }\bibfield  {title} {\enquote {\bibinfo {title} {{PESummary: the
  code agnostic Parameter Estimation Summary page builder}},}\ }\href {\doibase
  10.1016/j.softx.2021.100765} {\bibfield  {journal} {\bibinfo  {journal}
  {SoftwareX}\ }\textbf {\bibinfo {volume} {15}},\ \bibinfo {pages} {100765}
  (\bibinfo {year} {2021})},\ \Eprint {http://arxiv.org/abs/2006.06639}
  {arXiv:2006.06639 [astro-ph.IM]} \BibitemShut {NoStop}%
\bibitem [{\citenamefont {Gelman}(2013)}]{gelman2013bayesian}%
  \BibitemOpen
  \bibfield  {author} {\bibinfo {author} {\bibfnamefont {Carlin J. B. Stern H.
  S. Dunson D. B. Vehtari A. \& Rubin D.~B.}\ \bibnamefont {Gelman},
  \bibfnamefont {A.}},\ }\href@noop {} {\emph {\bibinfo {title} {Bayesian data
  analysis}}}\ (\bibinfo  {publisher} {CRC Press},\ \bibinfo {year}
  {2013})\BibitemShut {NoStop}%
\bibitem [{\citenamefont {Abbott}\ \emph
  {et~al.}(2021{\natexlab{c}})\citenamefont {Abbott} \emph
  {et~al.}}]{LIGOScientific:2020tif}%
  \BibitemOpen
  \bibfield  {author} {\bibinfo {author} {\bibfnamefont {R.}~\bibnamefont
  {Abbott}} \emph {et~al.} (\bibinfo {collaboration} {LIGO Scientific,
  Virgo}),\ }\bibfield  {title} {\enquote {\bibinfo {title} {{Tests of general
  relativity with binary black holes from the second LIGO-Virgo
  gravitational-wave transient catalog}},}\ }\href {\doibase
  10.1103/PhysRevD.103.122002} {\bibfield  {journal} {\bibinfo  {journal}
  {Phys. Rev. D}\ }\textbf {\bibinfo {volume} {103}},\ \bibinfo {pages}
  {122002} (\bibinfo {year} {2021}{\natexlab{c}})},\ \Eprint
  {http://arxiv.org/abs/2010.14529} {arXiv:2010.14529 [gr-qc]} \BibitemShut
  {NoStop}%
\end{thebibliography}%

\end{document}